%% file: manuscript.tex
\documentclass[trans]{IEEEtran}
\usepackage{dsfont}
\usepackage{amssymb}
\usepackage{subfig}
\usepackage{graphicx, amssymb, amsmath}
\usepackage{extarrows}
\usepackage{epstopdf}
\emergencystretch=\maxdimen
\IEEEoverridecommandlockouts
\begin{document}

\title{How Many Small Cells Can Be Turned off via Vertical Offloading under a Separation Architecture?}

\author{Shan~Zhang,~\IEEEmembership{Student~Member,~IEEE,}
        Jie~Gong,~\IEEEmembership{Member,~IEEE,}
        Sheng~Zhou,~\IEEEmembership{Member,~IEEE,}
        and~Zhisheng~Niu,~\IEEEmembership{Fellow,~IEEE}% <-this % stops a space
\thanks{The authors are with Tsinghua National Laboratory for Information Science and Technology, Tsinghua University, Beijing, 100084, P.R. China. Email: zhangshan11@mails.tsinghua.edu.cn, \{gongj13, sheng.zhou, niuzhs\}@tsinghua.edu.cn}% <-this % stops a space
\thanks{This work is sponsored in part by the National Basic Research Program of China (973 Program: 2012CB316001), the National Science Foundation of China (NSFC) under grant No. 61201191 and No. 61401250, the Creative Research Groups of NSFC under grant No. 61321061, and Hitachi R\&D Headquarter.}% <-this % stops a space
\thanks{Part of this work has been presented in Asilomar Conference on Signals, Systems, and Computers~2014 \cite{mine_asilomar}.}}

\maketitle

\begin{abstract}
To further improve the energy efficiency of heterogeneous networks, a separation architecture called hyper-cellular network (HCN) has been proposed, which decouples the control signaling and data transmission functions. Specifically, the control coverage is guaranteed by macro base stations (MBSs), whereas small cells (SCs) are only utilized for data transmission. Under HCN, SCs can be dynamically turned off when traffic load decreases for energy saving. A fundamental problem then arises: how many SCs can be turned off as traffic varies? In this paper, we address this problem in a theoretical way, where two sleeping schemes (i.e., random and repulsive schemes) with vertical inter-layer offloading are considered. Analytical results indicate the following facts: (1) Under the random scheme where SCs are turned off with certain probability, the expected ratio of sleeping SCs is inversely proportional to the traffic load of SC-layer and decreases linearly with the traffic load of MBS-layer; (2) The repulsive scheme, which only turns off the SCs close to MBSs, is less sensitive to the traffic variations; (3) deploying denser MBSs enables turning off more SCs, which may help to improve network energy-efficiency. Numerical results show that about 50\% SCs can be turned off on average under the predefined daily traffic profiles, and 10\% more SCs can be further turned off with inter-layer channel borrowing.
\end{abstract}

\section{Introduction}

    The multi-tier heterogeneous networks (HetNets), which consist of different types of base stations (BSs) (such as macro BSs, micro BSs, pico BSs and femto BSs), can effectively improve network capacity and thus are expected to be the dominant scenarios in the 5G era \cite{JSAC_overview_5G} \cite{5G_backhaul} \cite{NZhang_cloud_5G}.
    However, the huge energy consumption of HetNets brings heavy burdens to the network operators \cite{Overview_green} \cite{Green_MIMO}.
    Meanwhile, due to the dynamics of wireless traffic load, many BSs are lightly-loaded but still consume almost their peak energy on account of elements like airconditioner and power amplifier.
    Unfortunately, these low-efficient BSs can not be turned off for coverage guarantee, which makes the existing network energy inefficient \cite{Tango}.

\begin{figure}
    \center
    \includegraphics[width=3.5in]{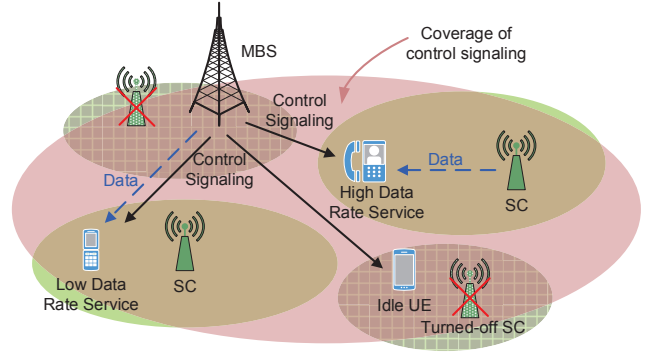}\\
    \caption{Network architecture of HCN.}\label{fig_architecture}
\end{figure}

    To solve this problem, we have proposed a new separation architecture called \emph{Hyper-Cellular Network} (HCN), whose main idea is to decouple the coverage of control signaling from the coverage of data transmission such that the data coverage can be more elastic in accordance with the traffic dynamics \cite{Hyper_cellular}.
    Under HCN, macro base stations (MBSs) and small cells\footnote{Small cells are the coverage of the BSs with relatively low transmit power (such as micro and pico BSs).} (SCs) play different roles as shown in Fig.~\ref{fig_architecture}.
    Generally, SCs are only utilized for high data rate transmission, whereas MBSs guarantee the network coverage and provide low data rate service.
    Therefore, each UE is always connected with the MBS-layer for control signaling, while their data traffics are served by MBSs or SCs depending on their service demands.
    For example, the UEs making video calls should keep dual connections with both MBSs and SCs, while UEs making voice calls are only served by MBSs.
    The HCN architecture can be realized by separating C/U planes \cite{Beyond_green}-\cite{Huawei}, which is a key technology for future 5G networks \cite{5G_Datang}\cite{5G_magzine}.

    With network coverage well-guaranteed by the MBS-layer, SCs can be turned off flexibly without causing coverage holes.
    At the same time, the quality of service (QoS) of the UEs within sleeping SCs can be satisfied by being offloaded either horizontally to neighboring or vertically to high-layer cells.
    Although offloading traffic from SCs to MBS increases the transmit power, turning off small cells saves energy as the constant power of SCs is much larger than the power consumed by data transmission \cite{EARTH}.
    Then a fundamental problem arises: how many SCs can be turned off via traffic offloading for given QoS requirements under HCN.

    As a starting point, we analyze the maximum ratio of sleeping SCs due to the traffic dynamics in time domain from the perspective of the whole network, where the UEs within sleeping SCs are only offloaded to MBSs (vertical offloading).
    In fact, the performance of horizontal offloading is limited due to the low transmit power of SCs, and accordingly a SC can only help to offload the traffic of neighboring cells within certain range.
    Therefore, horizontal offloading between SCs can not always be realized, and vertical offloading is sometimes the only choice for some SC to go into sleep if its neighboring SCs are not close enough.
    Besides, due to the dual connectivity of UEs in HCN, vertical offloading can be easily implemented without handovers.
    Two sleeping schemes are considered in this paper:
    \begin{enumerate}
      \item Random scheme: each SC is turned off with equivalent probability $p_\mathrm{s}$;
      \item Repulsive scheme: the SCs whose distance to the nearest MBSs smaller than $R_\mathrm{s}$ go into sleep, whereas the other SCs remain active.
    \end{enumerate}
    Then, our problem is to find the maximum $p_\mathrm{s}$ and $R_\mathrm{s}$ which satisfy the given outage constraints.
    To conduct theoretical analysis, a two-layer HCN is considered, where MBSs are assumed to be regularly deployed as hexagonal cells whereas the distribution of SCs is modeled as Voronoi tessellation of a homogeneous Poisson Point Process (PPP).
    The main contributions of this paper include:
    \begin{itemize}
      \item The approximated closed-form expressions of the outage probability are derived under the two sleeping schemes, which are validated through extensive simulations.
      \item The maximum sleeping ratio of SCs under the two scheme is derived based on the analytical outage probability, and the influences of system parameters (such as traffic load and BS density) are analyzed.
      \item We also consider the case when the redundant bandwidth of the SC-layer can be released and re-allocated to the MBS-layer to serve more offloaded UEs (channel borrowing).
      \item Under a typical scenario and two different daily traffic profiles, we show that about 50\% SCs can go into sleep on average, and 10\% more SCs can be further turned off if channel borrowing is conducted.
    \end{itemize}

    In summary, the analytical results of the maximum sleeping ratio of SCs are obtained under the random and repulsive schemes, which offers a guideline for real network planning and operations. In addition to the implementation considerations, the sleeping algorithm design of the separation architecture is consistent with that of the conventional networks. Therefore, our method also applies to the conventional two-tier heterogeneous networks.

    The rest of this paper is organized as follows.
    Related work is introduced in Section~\ref{sec_related_work}.
    System model is described in Section~\ref{sec_sys_model}, and the outage probability is derived in Section~\ref{sec_outage_probability}.
    Then, the random and repulsive schemes are analyzed in Section~\ref{sec_schemes} and the two schemes are evaluated under daily traffic profiles in Section~\ref{sec_traffic}.
    At last, Section~\ref{sec_conclusions} concludes this paper.

%%%%%%%%%%%%%%%%%%%%%%%%%%%%%%%%%%%%%%%%%%%%%%%%%%%%%%%%%%%%%%%%%%%%%%%%%%%%%%%%%%%%%%%%%%%%%%%%%%%
\section{Related Work}
    \label{sec_related_work}
    \input{RelatedWork.tex}
%%%%%%%%%%%%%%%%%%%%%%%%%%%%%%%%%%%%%%%%%%%%%%%%%%%%%%%%%%%%%%%%%%%%%%%%%%%%%%%%%%%%%%%%%%%%%%%%%%%
\section{System Model}
    \label{sec_sys_model}
    \input{SystemModel.tex}
%%%%%%%%%%%%%%%%%%%%%%%%%%%%%%%%%%%%%%%%%%%%%%%%%%%%%%%%%%%%%%%%%%%%%%%%%%%%%%%%%%%%%%%%%%%%%%%%%%%
\section{Outage Probability Analysis}
    \label{sec_outage_probability}
    \input{OutageProbability.tex}
%%%%%%%%%%%%%%%%%%%%%%%%%%%%%%%%%%%%%%%%%%%%%%%%%%%%%%%%%%%%%%%%%%%%%%%%%%%%%%%%%%%%%%%%%%%%%%%%%%%
\section{Problem Analysis and Solutions}
    \label{sec_schemes}
    \input{ProblemSolution.tex}

%%%%%%%%%%%%%%%%%%%%%%%%%%%%%%%%%%%%%%%%%%%%%%%%%%%%%%%%%%%%%%%%%%%%%%%%%%%%%%%%%%%%%%%%%%%%%%%%%%%
\section{Performance Evaluation under Daily Traffic Profiles}
    \label{sec_traffic}
    \input{TrafficProfiles.tex}
%%%%%%%%%%%%%%%%%%%%%%%%%%%%%%%%%%%%%%%%%%%%%%%%%%%%%%%%%%%%%%%%%%%%%%%%%%%%%%%%%%%%%%%%%%%%%%%%%%%
\section{Conclusions}
    \label{sec_conclusions}
    In this paper, the expected sleeping ratio of SCs with the time-varying traffic is obtained for two SC sleeping schemes (random and repulsive schemes), under which the UEs of the sleeping SCs are vertically offloaded to the MBS-layer for outage probability guarantee.
    Under the random scheme, the ratio of sleeping SCs is inversely proportional to the density of SC UEs whereas decreases linearly with the density of MBS UEs.
    Compared with the random scheme, the repulsive scheme performs better when the traffic load exceeds certain thresholds.
    In addition, the analytical results suggest that deploying more MBSs can help to save energy by offering more opportunities for SC sleeping.
    Furthermore, if the spectrum resource can be dynamically borrowed between layers, more SCs can be turned off especially when the MBS-layer is heavily-loaded.
    Numerical results show that half of the SCs can be turned off on average under two typical daily traffic profiles, and 10\% more SCs can further go into sleep if channel borrowing is allowed.
    For future work, the energy-optimal network planning should be analyzed based on the obtained sleeping ratio.
    In addition, detailed energy-efficient SC sleeping schemes should be designed, where the vertical and horizontal offloading schemes can be jointly optimized, and more realistic network scenarios (such as non-uniform traffic load) should be considered.

%%%%%%%%%%%%%%%%%%%%%%%%%%%%%%%%%%%%%%%%%%%%%%%%%%%%%%%%%%%%%%%%%%%%%%%%%%%%%%%%%%%%%%%%%%%%%%%%%%%

\appendices{}
   \input{appendix.tex}

\begin{IEEEbiography}[{\includegraphics[width=1in,height=1.25in,clip,keepaspectratio]{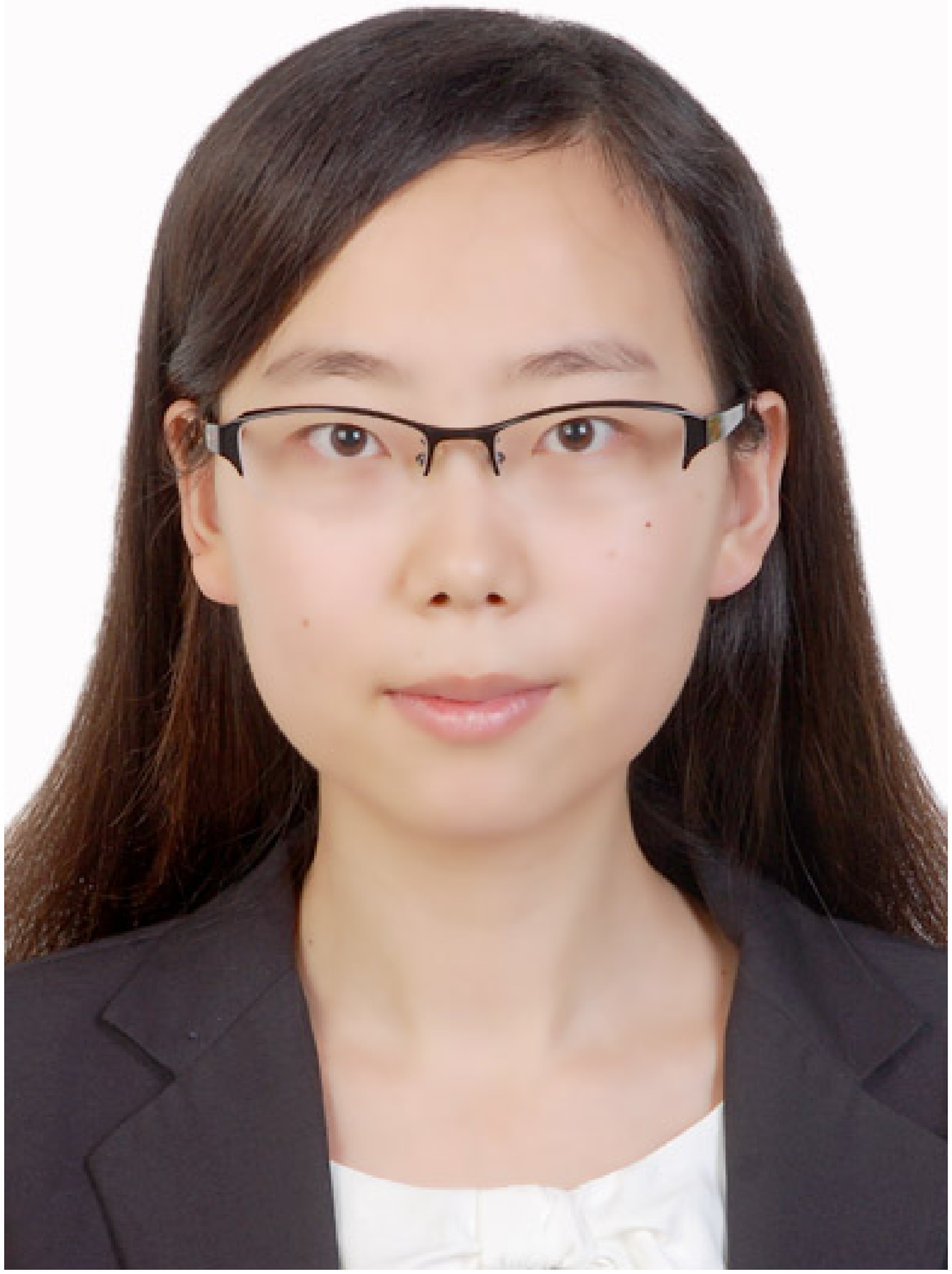}}]{Shan Zhang} 
received B.S. degree in Electronic Engineering from Beijing Institute Technology, Beijing, China, in 2011 and is currently a Ph.D. candidate in Department of Electronic Engineering, Tsinghua University. She received the Best Paper Award from the 19th Asia-Pacific Conference on Communication (APCC) in 2013. Her research interests include network planning, resource and traffic management for green communications.
\end{IEEEbiography}
\vspace{-20mm}
\begin{IEEEbiography}[{\includegraphics[width=1in,height=1.25in,clip,keepaspectratio]{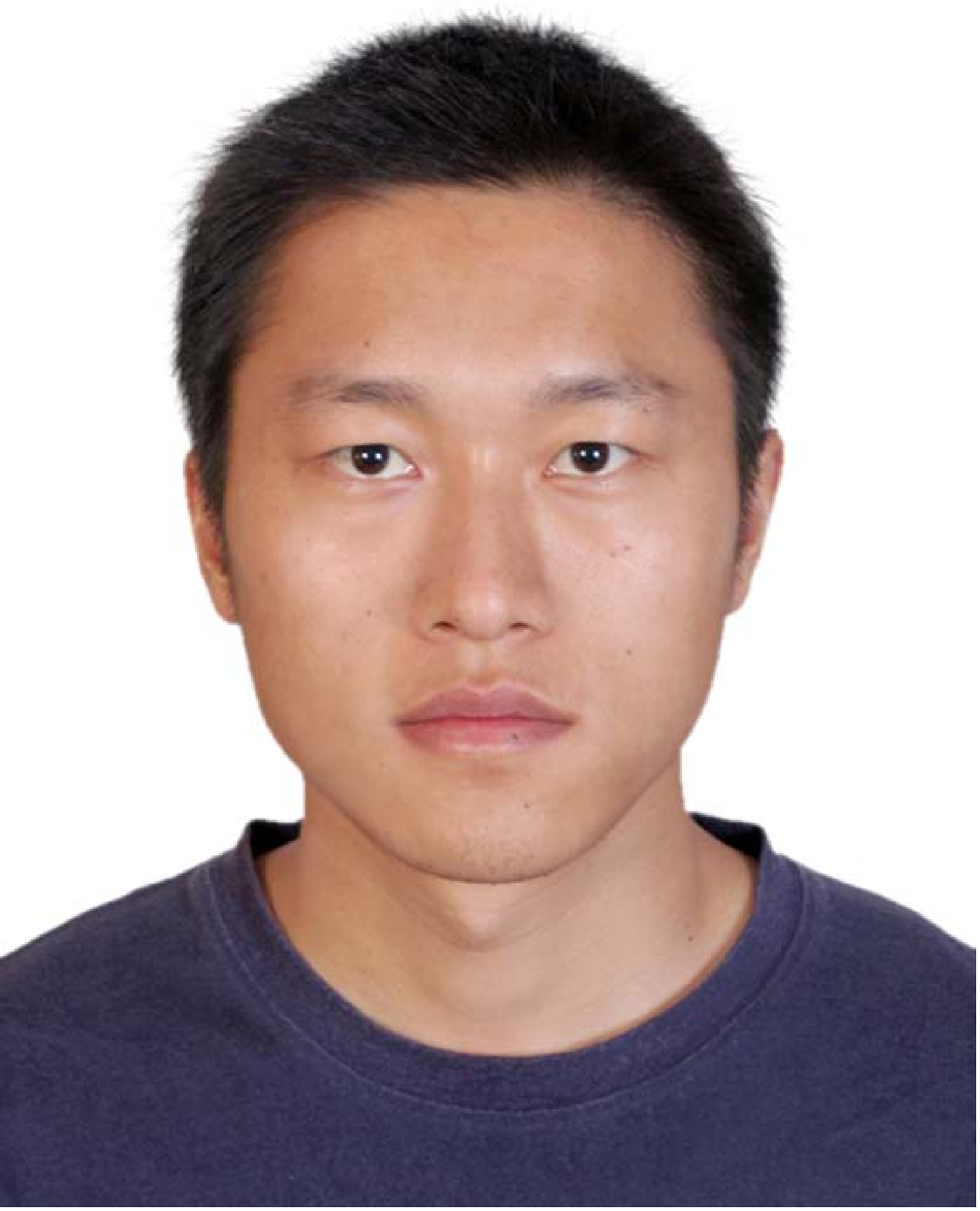}}]{Jie Gong}
received the B.S. and Ph.D. degrees from Tsinghua University, Beijing, China, in 2008 and 2013, respectively. He is currently a Postdoctoral Scholar with the Department of Electronic Engineering, Tsinghua University. From July 2012 to January 2013, he visited the Institute for Digital Communications, The University of Edinburgh, Edinburgh, U.K. His research interests include base station cooperation in cellular networks, energy harvesting, and green wireless communications. Dr. Gong was a co-recipient of the Best Paper Award from the IEEE Communications Society Asia Pacific Board in 2013.
\end{IEEEbiography}
\vspace{-20mm}
\begin{IEEEbiography}[{\includegraphics[width=1in,height=1.25in,clip,keepaspectratio]{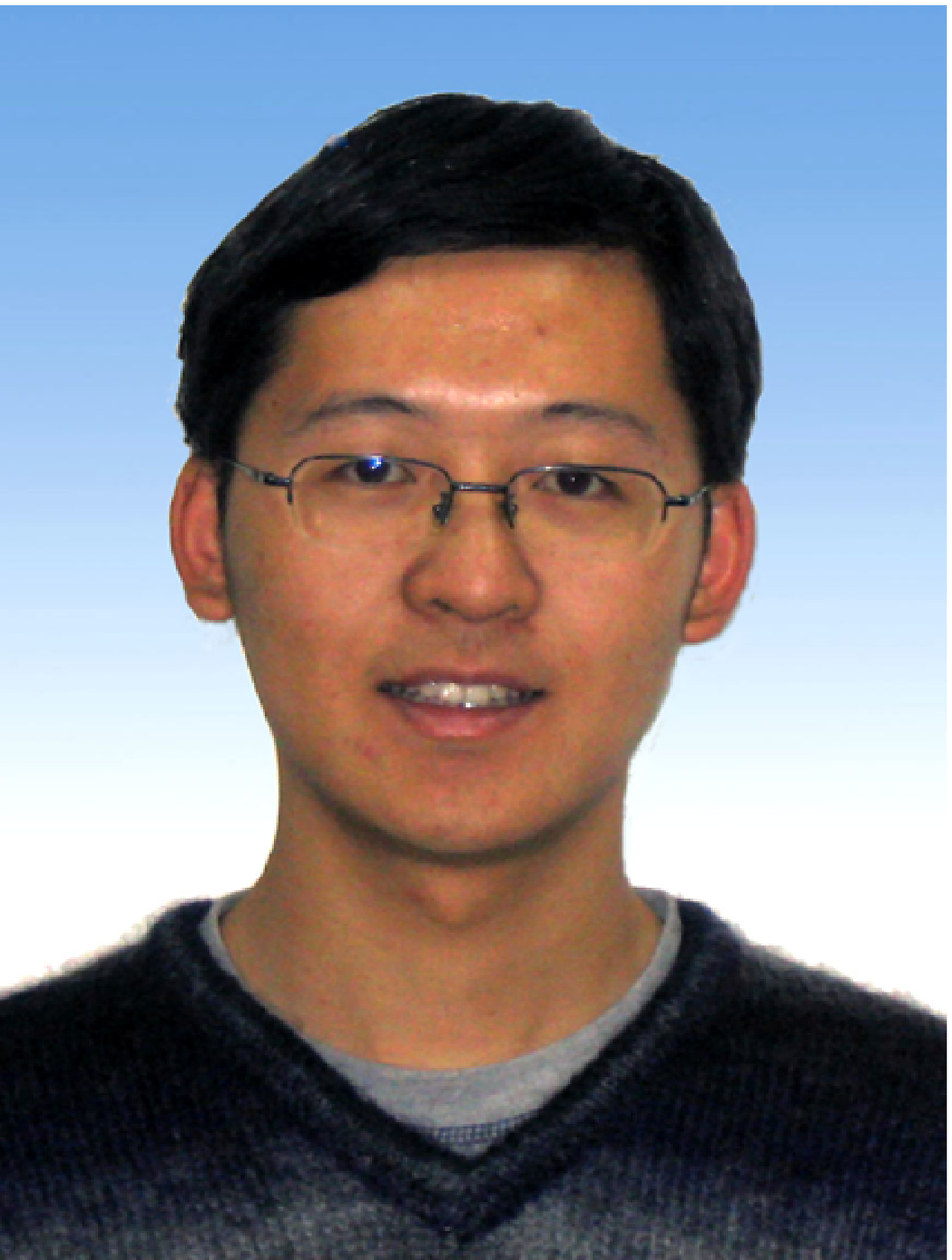}}]{Sheng Zhou}
(S'06, M'12) received his B.S. and Ph.D. degrees in Electronic Engineering from Tsinghua University, China, in 2005 and 2011, respectively. He is now a postdoctoral scholar in Electronic Engineering Department at Tsinghua University, Beijing, China. From January to June 2010, he was a visiting student at Wireless System Lab, Electrical Engineering Department, Stanford University, CA, USA. He is a co-recipient of the Best Paper Award from the 15th Asia-Pacific Conference on Communication (APCC) in 2009, and the 23th IEEE International Conference on Communication Technology (ICCT) in 2011.  His research interests include cross-layer design for multiple antenna systems, cooperative transmission in cellular systems, and green wireless cellular communications.
\end{IEEEbiography}
\vspace{-20mm}
\begin{IEEEbiography}[{\includegraphics[width=1in,height=1.25in,clip,keepaspectratio]{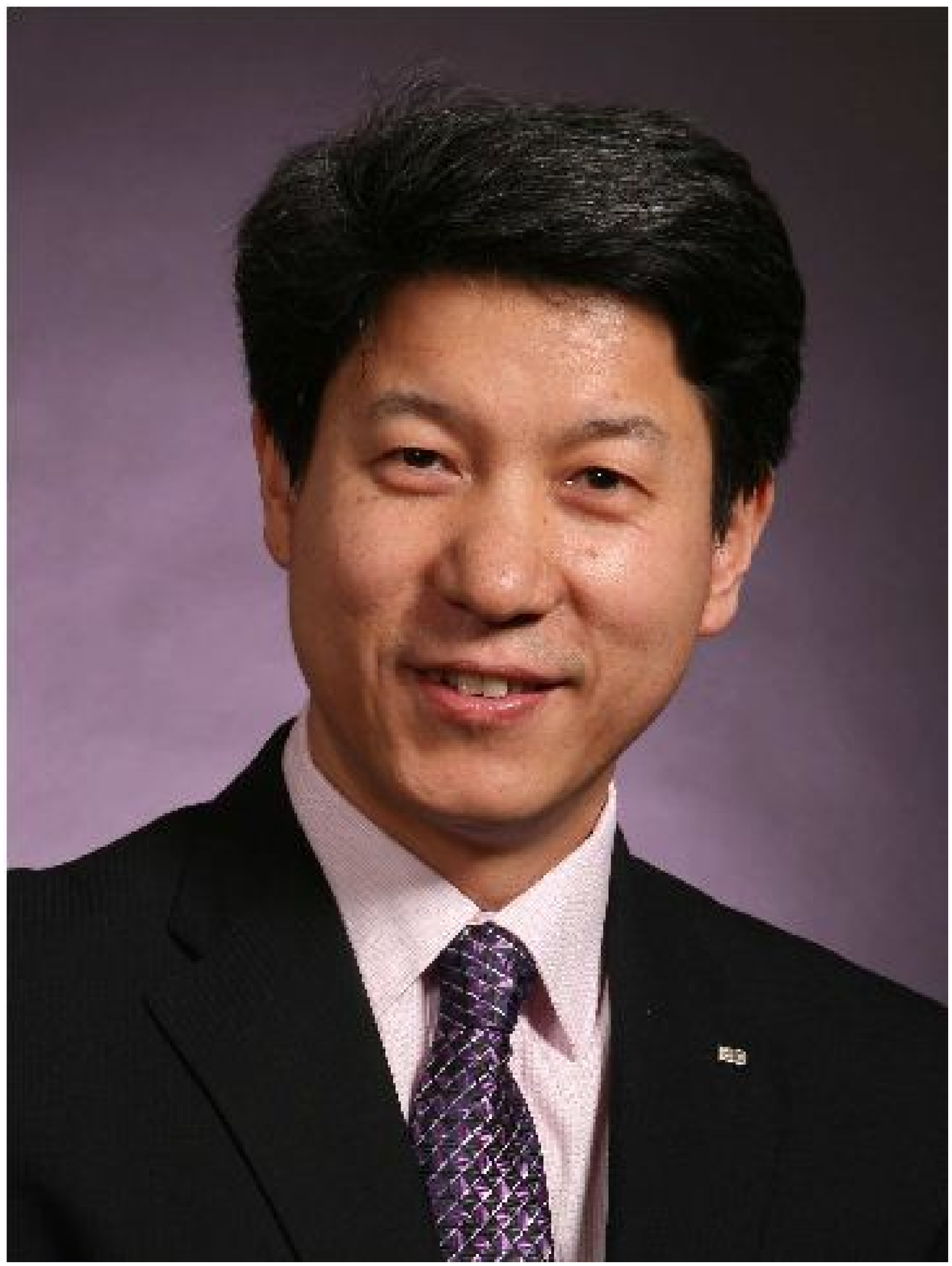}}]{Zhisheng Niu}
(M'98-SM'99-F'12) graduated from Northern Jiaotong University (currently Beijing Jiaotong University), Beijing, China, in 1985, and got his M.E. and D.E. degrees from Toyohashi University of Technology, Toyohashi, Japan, in 1989 and 1992, respectively.  After spending two years at Fujitsu Laboratories Ltd., Kawasaki, Japan, he joined with Tsinghua University, Beijing, China, in 1994, where he is now a professor at the Department of Electronic Engineering, deputy dean of the School of Information Science and Technology, and director of Tsinghua-Hitachi Joint Lab on Environmental Harmonious ICT.  He is also a guest chair professor of Shandong University.  His major research interests include queueing theory, traffic engineering, mobile Internet, radio resource management of wireless networks, and green communication and networks.
Dr. Niu has been an active volunteer for various academic societies, including Director for Conference Publications (2010-11) and Director for Asia-Pacific Board (2008-09) of IEEE Communication Society, Membership Development Coordinator (2009-10) of IEEE Region 10, Councilor of IEICE-Japan (2009-11), and council member of Chinese Institute of Electronics (2006-11).  He is now a distinguished lecturer (2012-13) of IEEE Communication Society, standing committee member of both Communication Science and Technology Committee under the Ministry of Industry and Information Technology of China and Chinese Institute of Communications (CIC), vice chair of the Information and Communication Network Committee of CIC, editor of IEEE Wireless Communication Magazine.
Dr. Niu received the Outstanding Young Researcher Award from Natural Science Foundation of China in 2009 and the Best Paper Awards (with his students) from the 13th and 15th Asia-Pacific Conference on Communication (APCC) in 2007 and 2009, respectively.  He is now a fellow of both IEEE and IEICE.
\end{IEEEbiography}
\vspace{-20mm}

\end{document}

%% file: RelatedWork.tex
    Under the traditional network architecture, BSs can not be turned off flexibly due to the requirement of coverage guarantee.
    To tackle this problem, \emph{cell zooming} is proposed to compensate for the sleeping cells by enlarging the coverage of neighboring BSs, which can be realized through power control and antenna titling \cite{Cell_zooming}.
    Meanwhile, \emph{horizontal offloading} is widely adopted for QoS guarantee \cite{SZhou_BS_sleeping}-\cite{Tony_single_tier_sleep}, under which the UEs of sleeping cells are offloaded to neighboring cells regardless of their cell types.
    SC sleeping through horizontal offloading under HetNets is investigated theoretically in \cite{DCao}-\cite{Repulsive}.
    In \cite{DCao}, the upper and lower bounds of the optimal density of active SCs are derived under time-varying traffic load, where SCs are turned off randomly.
    Instead, the SCs close to MBSs are turned off in \cite{Repulsive}, after which the corresponding UEs will be re-associated to active cells based on the received signal strength.
    However, coverage holes may still exist via these horizontal offloading schemes in real systems due to the complex wireless environment.
    Therefore, \emph{vertical offloading} is adopted as an alternative solution to guarantee QoS in the multi-tier HetNets \cite{forcast_exponential_smoothing}-\cite{vertical_MDP}, where the UEs of sleeping SCs are offloaded to MBSs instead of neighboring SCs.
    In \cite{forcast_exponential_smoothing}-\cite{forcast_ANN}, SCs are dynamically turned on/off based on the predicted future traffic load.
    Besides, a dynamic energy-optimal SC sleeping algorithm is proposed based on Markov Decision Process \cite{vertical_MDP}.
    Whereas these studies have high complexity and are limited to small scale networks.

    Recently, separation architecture has received more attention.
    A SC sleeping scheme under HCN is proposed and optimized in \cite{Mine_Globecom_load}, where the SCs are turned off probabilistically according to their traffic load via vertical offloading.
    Furthermore, the energy saving gain of HCN is firstly analyzed in \cite{WZhang_separation}, where random sleeping scheme via horizontal offloading are considered.
    The main differences between \cite{WZhang_separation} and our work are that we focus on vertical offloading, which has been rarely addressed under the separation architecture.

%% file: SystemModel.tex
\begin{figure}
    \center
    \includegraphics[width=3in]{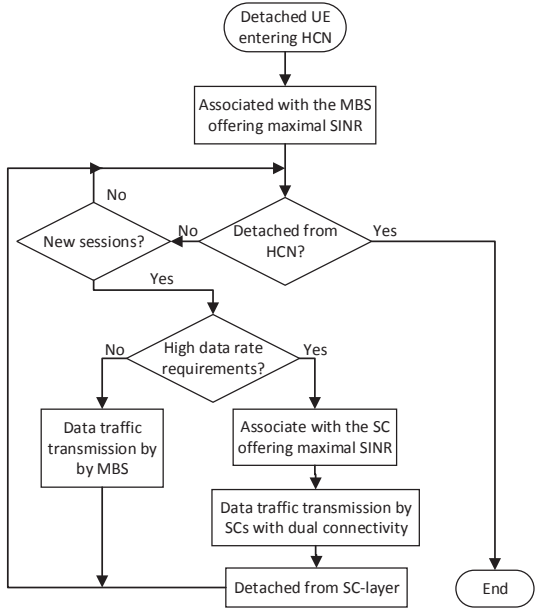}\\
    \caption{Service process for a typical UE under HCN.}\label{fig_UE_association}
\end{figure}

    We consider a two-layer HCN, under which the service procedure for a UE is described in Fig.~\ref{fig_UE_association}.
    When a detached UE arrives, it will firstly connect to the MBS-layer for basic control signaling, and the MBS which offers maximal average Signal to Interference and Noise Ratio (SINR) will be selected.
    Then, the UE will always connect the MBS-layer until it leaves the HCN.
    During this period, the UE initiated sessions will be served according to the data rate demand and mobility.
    The SCs are utilized for high data rate transmission (like real-time video, online game), whereas MBSs mainly guarantee lower data rate services (such as voice).
    Besides, high mobile UEs will be served by MBSs to avoid frequent handovers.
    Therefore, only the UEs with low mobility and high data rate requirement will choose a SC for data service, during which they maintain dual connections with both layers for signaling and data respectively.
    Notice that their connections with the SC-layer will be dropped once their sessions are completed.

    According to the service status, all active UEs in HCN can be classified into 2 classes:
    \begin{enumerate}
      \item \emph{MBS UEs}: UEs served by MBSs with low data rate;
      \item \emph{SC UEs}: UEs served by SCs with high data rate.
    \end{enumerate}
    Because we focus on the time dynamics of traffic load instead of the spatial non-uniformity, we assume the distributions of MBS UEs and SC UEs both  follow homogeneous PPPs with different densities.

\begin{figure}
    \center
    \includegraphics[width=2in]{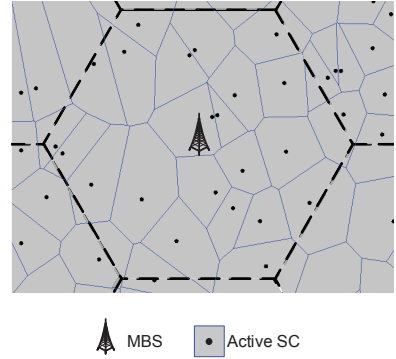}\\
    \caption{Network topology.}\label{fig_topology}
\end{figure}

    As for the network topology, the MBSs are assumed to be regularly deployed as hexagonal cells whereas SCs are assumed to be the Voronoi tessellation of a homogeneous PPP process as shown in Fig.~\ref{fig_topology}, due to the different roles of MBSs and SCs.
    Recall that MBSs are expected to guarantee the network coverage, therefore their locations should be carefully designed.
    On the contrary, SCs are deployed only to boost the network capacity, whose locations can be quite random.
    In fact, the traditional regular hexagonal cells and the PPP have similar accuracy when used to model the distribution of BSs in real systems, whereas the former model is the ideal case and the latter offers a performance lower bound \cite{JAndrews}.

\subsection{Bandwidth Allocation}

    Orthogonal bandwidth is used by different layers to avoid the severe inter-layer interference, especially for protecting the signaling coverage.
    Furthermore, the spectrum reuse factor within each layer is set to 1.
    When some SCs are turned off with their UEs offloaded to MBSs, the spectrum resource of the MBS-layer is further divided into two parts to serve their associated MBS UEs and the offloaded SC UEs respectively.
    In addition, more SCs can be turned off if the redundant bandwidth of the SC-layer can be released and reallocated to the MBS-layer for traffic offloading (named as \emph{Channel Borrowing (CB)}).
    Note that the borrowed bandwidth cannot be used by the SC-layer.
    In other words, the bandwidth used by the offloaded UEs is given by
    \begin{equation}\label{eq_bandwidth_offloaded}\footnotesize
        w_\mathrm{o} = \left\{ \begin{array}{ll} W_\mathrm{m}+W_\mathrm{s} - w_\mathrm{m}- w_\mathrm{s}, & \mbox{with CB} \\ W_\mathrm{m} - w_\mathrm{m}, & \mbox{no CB} \end{array},\right.
    \end{equation}
    where $W_\mathrm{m}$ and $W_\mathrm{s}$ are the total bandwidth initially pre-allocated to the MBS-layer and SC-layer, $w_\mathrm{m}$ and $w_\mathrm{s}$ are the bandwidth used by the non-offloaded UEs at the two layers respectively.
    Obviously, CB offers more opportunity for SC sleeping by making better use of the spectrum resource.

\subsection{Link Layer model}

        Assume all MBSs transmit at a fixed power $P^\mathrm{m}$.
        For a typical UE$_u$, the received SINR is given as follows if it is served by MBS$_i$.
            \begin{equation}\label{eq_SINR_MBS}\footnotesize
                \gamma^{\mathrm{m}}_{iu} = \frac{P^\mathrm{m} (d^\mathrm{m}_{iu})^{-\alpha_\mathrm{m}} h^\mathrm{m}_{iu}} { \sum\limits_{j\in \mathcal{B}^\mathrm{m},j \neq i} P^\mathrm{m} (d^\mathrm{m}_{ju})^{-\alpha_\mathrm{m}} h^\mathrm{m}_{ju} + \sigma^2 },
            \end{equation}
        where $\sigma^2$ is the noise power, $\mathcal{B}^\mathrm{m}$ is the set of active MBSs in the network, $d^\mathrm{m}_{iu}$ is the distance between UE$_u$ and MBS$_i$, $\alpha_\mathrm{m}$ is the path loss factor of MBS-layer, and $h^\mathrm{m}_{iu}$ is an exponential random variable with mean 1 incorporating the effect of Rayleigh fading.
        Assume each BS allocates resource (e.g., time slots or wireless spectrum) equally to its associated active UEs, then the achievable rate of UE$_u$ is given by:
            \begin{equation}\label{eq_r_d}\footnotesize
                r^\mathrm{m}_{iu} = \frac{w_\mathrm{m}}{N_{i}^\mathrm{m}+1} \log_2 (1 + \gamma^\mathrm{m}_{iu}),
            \end{equation}
        where $w_\mathrm{m}$ is the bandwidth used by each MBS for its associated UEs, and $N_{i}^\mathrm{m}$ is a random variable denoting the number of active residual UEs served by MBS$_i$ except UE$_u$.
        Then the outage probability that the rate of UE$_{u}$ is less than the predefined threshold $U_\mathrm{m}$ is given by $\mathds{P}\left\{r^\mathrm{m}_{iu}<U_\mathrm{m}\right\}$.
        By averaging this probability over the possible position of UE$_{u}$, and $N_{i}^\mathrm{m}$, we can get the service outage constraint of MBS UEs as\footnote{The subscripts $i$ and $u$ are omitted here for simplicity.}:
        \begin{equation}\label{eq_outage_MBS}\footnotesize
            G_\mathrm{m} = \mathbb{E}_{\{N_\mathrm{m}, d_\mathrm{m}\}} \left\{ \mathds{P} \left( \frac{w_\mathrm{m}}{N_\mathrm{m}+1} \log_2 (1+\gamma_\mathrm{m}) < U_\mathrm{m} \Big| N_\mathrm{m}, d_\mathrm{m} \right) \right\} < \eta_\mathrm{m}.
        \end{equation}

        Assume all SBSs also adopt the constant transmit power $P^\mathrm{s}$.
        If UE$_u$ is served by SC$_k$, its received SINR is given by:
            \begin{equation}\label{eq_SINR_MBS}\footnotesize
                \gamma^{\mathrm{s}}_{ku} = \frac{P^\mathrm{s} (d^\mathrm{s}_{ku})^{-\alpha_\mathrm{s}} h^\mathrm{s}_{ku}} { \sum\limits_{l\in \mathcal{B}^\mathrm{s},l \neq k} P^\mathrm{s} (d^\mathrm{s}_{lu})^{-\alpha_\mathrm{s}} h^\mathrm{s}_{ku} + \sigma^2 },
            \end{equation}
        where $\mathcal{B}^\mathrm{s}$ is the set of active SCs in the network, $d^\mathrm{s}_{ku}$ is the distance between UE$_u$ and SC$_k$, $\alpha_\mathrm{s}$ is the path loss factor of SC-layer, and $h^\mathrm{s}_{ku}$ is an exponential random variable with mean 1 incorporating the effect of Rayleigh fading.
        Similarly, the outage probability constraint of the SC UEs can be obtained\footnote{The subscripts $k$ and $u$ are omitted here for simplicity.}:
         \begin{equation}\label{eq_outage_PBS}\footnotesize
            G_\mathrm{s} = \mathbb{E}_{\{A_\mathrm{s}, N_\mathrm{s}, d_\mathrm{s}\}} \left\{ \mathds{P} \left( \frac{w_\mathrm{s}}{N_\mathrm{s}+1} \log_2 (1+\gamma_\mathrm{s}) < U_\mathrm{s} \Big| N_\mathrm{s}, d_\mathrm{s}, A_\mathrm{s} \right) \right\} < \eta_\mathrm{s},
        \end{equation}
        where $w_\mathrm{s}$ is the bandwidth used by each SC, $N_\mathrm{s}$ is the number of residual UEs in the target SC, $\gamma_\mathrm{s}$ is the spectrum efficiency of the considered SC UE, whose distance to the target SC is denoted as $d_\mathrm{s}$.
        In addition, the cell size of the target SC is also random, denoted as $A_\mathrm{s}$.

        As for a typical offloaded UE, its outage probability can be derived in the same way:
        \begin{equation}\label{eq_outage_offloaded}\footnotesize
            G_\mathrm{o} = \mathbb{E}_{\{N_\mathrm{o}, d_\mathrm{o}\}} \left\{ \mathds{P} \left( \frac{w_\mathrm{o}}{N_\mathrm{o}+1} \log_2 (1+\gamma_\mathrm{o}) < U_\mathrm{o} \Big| N_\mathrm{o}, d_\mathrm{o} \right) \right\} < \eta_\mathrm{o},
        \end{equation}
        where $d_\mathrm{o}$ is the distance between the typical offloaded UE to its associated MBS, $N_o$ denotes the number of remaining offloaded UEs in the same MBS cell, and $\gamma_\mathrm{o}$ means the received SINR varying with channel fading.
        Notice that the QoS of the offloaded UEs will not degrade if $U_\mathrm{o}=U_\mathrm{s}$ and $\eta_\mathrm{o}=\eta_\mathrm{s}$.

    \begin{figure}[!t]
        \centering
        \includegraphics[width=2in]{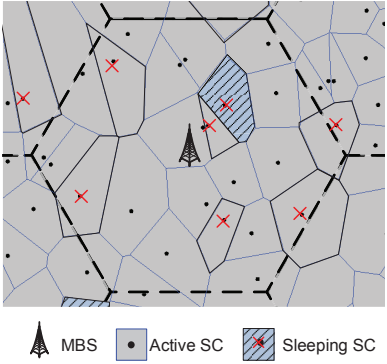}
        \caption{Illustration of the random scheme.}
        \label{fig_sleeping_rand}
    \end{figure}

\subsection{SC Sleeping Schemes}

    Our problem can be formulated to maximize the ratio of SCs turned off under the outage probability constrains of Eqs.~(\ref{eq_outage_MBS}), (\ref{eq_outage_PBS}) and (\ref{eq_outage_offloaded}).
    However, this problem is hard to solve as there exist no closed-form expressions of the outage probability.
    In fact, even the distributions of $N_\mathrm{s}$, $N_\mathrm{o}$, $A_\mathrm{s}$, and $d_\mathrm{o}$ do not have general expressions.

\begin{figure}[!t]
        \centering
        \includegraphics[width=2in]{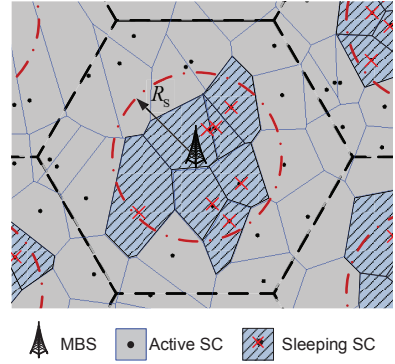}
        \caption{Illustration of the repulsive scheme.}
        \label{fig_sleeping_repulsive}
    \end{figure}

    To conduct theoretical analysis, we consider two basic sleeping schemes: random scheme and repulsive scheme.
    Under the random scheme, SCs are treated equivalently and go into sleep independently with probability $p_\mathrm{s}$.
    Whereas, SCs are differentiated by their distance to the MBSs under the repulsive scheme, and only the SCs whose distance to the nearest MBSs is less than sleeping radius $R_\mathrm{s}$ can be turned off.
    The two schemes are illustrated in Fig.~\ref{fig_sleeping_rand} and Fig.~\ref{fig_sleeping_repulsive} respectively.

    The random scheme has been adopted in many studies, which is considered as a a baseline for cell sleeping \cite{Tony_single_tier_sleep}, \cite{DCao}, and \cite{WZhang_separation}.
    Besides, the random sleeping can be easily implemented as it requires no extra information like traffic load and locations of each cell, and the average ratio of sleeping SCs is equal to the sleeping probability $p_\mathrm{s}$.
    Under the repulsive scheme, the offloaded UEs generally enjoy smaller path loss and larger spectrum efficiency for shorter access distance to the MBSs, and thus the QoS of the offloaded UEs can be more easily guaranteed.
    In this case, the average sleeping ratio of SCs is given by $\frac{\pi R_\mathrm{s}^2}{\frac{3\sqrt{3}}{2}D^2}$, where $D$ is the coverage radius of each MBS cell.
    
    When power control is not implemented, the power consumption of a BS becomes a constant \cite{EARTH}. Thus, the network power consumption can be treated as the weighted sum of the MBS and SC densities. Therefore, the ratio of sleeping SCs reflects the energy saving gain.

    %\begin{figure}
    %\center
    %\includegraphics[width=3.5in]{CA.eps}\\
    %\caption{Bandwidth Allocation with/without CA}\label{fig_CA}
    %\end{figure}

%% file: OutageProbability.tex
    \subsection{Outage Constraint of MBS UEs}

    For the outage probability of MBS UEs given by Eq.~(\ref{eq_outage_MBS}), we have
    \begin{equation}\label{eq_blocking_pr_MBS} \footnotesize
        \begin{split}
            & G_\mathrm{m} \!=\! \mathbb{E}_{\{N_\mathrm{m}, d_\mathrm{m}\}} \left\{\mathds{P}\left\{ \frac{w_\mathrm{m}}{N_\mathrm{m}+1} \log_2(1+\gamma_\mathrm{m})<U_\mathrm{m} \Big| N_\mathrm{m}, d_\mathrm{m} \right\}\right\} \\
            & \!=\! \int\limits_{0}\limits^{D} \sum\limits_{n=0}\limits^{\infty} \mathds{P} \left\{ \gamma_\mathrm{m}\! < \!2^{\frac{(n+1)U_\mathrm{m}}{w_\mathrm{m}}}\!-\!1 \Big| d\right\} p_{N_\mathrm{m}}(n) f_{d_\mathrm{m}}(d) \mathrm{d} d ,
        \end{split}
    \end{equation}
    where $p_{N_\mathrm{m}}(n)$ is the probability that the target MBS has $n$ residential MBS UEs except UE$_u$, and $f_{d_\mathrm{m}}(d)$ is the probability density function of $d_\mathrm{m}$.
    As the distribution of MBS UEs follows homogeneous PPP process, $N_\mathrm{m}$ follows the Poisson distribution of parameter $\frac{3\sqrt{3}}{2}\lambda_\mathrm{m} D^2 $ according to Slivnyak-Mecke theorem \cite{Stochastic_Geometry}, where $\lambda_\mathrm{m}$ is the density of MBS UEs.
    In addition, $f_{d_\mathrm{m}}(d)=\frac{2\pi }{D} d$ by assuming UEs to be uniformly distributed within a circle cell of radius $D$.
    The closed-form expression of $G_\mathrm{m}$ can be derived when the received SNR is high, which generally holds for current cellular systems.

    \textbf{Theorem~1.} As $\frac{\sigma^2}{P^\mathrm{m}}\rightarrow0$, the outage probability of MBS UEs is given by
        \begin{equation}\label{eq_outage_MBS_1}\footnotesize
            G_\mathrm{m} \!=\! \frac{2 D^{\alpha_\mathrm{m}} (I\!+\!1)\sigma^2}{P^\mathrm{m} \left(\alpha_\mathrm{m}\!+\!2\right) } \! \left(2^{\frac{U_\mathrm{m}}{w_\mathrm{m}}}\exp\left(\frac{3\sqrt{3}}{2} D^2 \lambda_{\mathrm{m}} \left( 2^{\frac{U_\mathrm{m}}{w_\mathrm{m}}} \!-\! 1 \right)\right)\!-\!1\right),
        \end{equation}
    where $I$ denotes the ratio of inter-cell interference to noise.
    \emph{Proof}: See Appendix~\ref{appendix_CBS}.
    \hfill \rule{4pt}{8pt}\\

    Furthermore, when $\frac{U_\mathrm{m}}{w_\mathrm{m}} \rightarrow 0$, Eq.~(\ref{eq_outage_MBS_1}) can be further simplified:
        \begin{equation}\label{eq_outage_MBS_2}\footnotesize
            G_\mathrm{m} \!=\! \frac{2 D^{\alpha_\mathrm{m}} (I\!+\!1)\sigma^2}{P^\mathrm{m} \left(\alpha_\mathrm{m}\!+\!2\right) } \! \left(2^{\frac{U_\mathrm{m}}{w_\mathrm{m}}\left(1 + \frac{3\sqrt{3}}{2} D^2 \lambda_{\mathrm{m}} \frac{U_\mathrm{m}}{w_\mathrm{m}} \right)}\!-\!1\right),
        \end{equation}
    and the service outage constraint of MBS UEs Eq.~(\ref{eq_outage_MBS}) is equivalent to
        \begin{equation}\label{eq_outage_MBS_simple}\footnotesize
            \bar{w}_\mathrm{m} \log_2 \left( 1 + \tau_\mathrm{m} \right) \geq U_\mathrm{m},
        \end{equation}
    where $\bar{w}_\mathrm{m}=\frac{w_\mathrm{m}}{1+ \frac{3\sqrt{3}}{2}\lambda_\mathrm{m} D^2}$ is the expected bandwidth allocated to each MBS UE, and $\tau_\mathrm{m}$ denotes the received SINR of cell edge UEs depending on $\eta_\mathrm{m}$ given by
        \begin{equation}\label{eq_tau_MBS}\footnotesize
            \tau_\mathrm{m} = \frac{P^\mathrm{m}}{\sigma^2(I+1)} \frac{\alpha_\mathrm{m}+2}{2} \frac{\eta_\mathrm{m}}{D^{\alpha_\mathrm{m}}}.
        \end{equation}
    Notice that Eq.~(\ref{eq_outage_MBS_simple}) is a linear constraint on $w_\mathrm{m}$, and its physical meaning is that the average data rate of the non-cell-edge UEs (received SINR above $\tau_\mathrm{m}$) should be no smaller than the $U_\mathrm{m}$.

    The physical meaning of $\frac{U_\mathrm{m}}{w_\mathrm{m}} \rightarrow 0$ is that the data rate requirement is relatively low compared with the spectrum resource, which is reasonable as each MBS usually supports large number of UEs simultaneously in real systems.
    Therefore, Eq.~(\ref{eq_outage_MBS_simple}) can be applied to simplify the service outage constraint of MBS UEs.

    \subsection{Outage Constraint of SC UEs}

    The outage probability for a typical SC UE (Eq.~(\ref{eq_outage_PBS})) is given by
    \begin{equation}\label{eq_blocking_pr_PBS}\footnotesize
        \begin{split}
            & G_\mathrm{s} \!=\! \mathbb{E}_{\{A_\mathrm{s}, N_\mathrm{s}, d_\mathrm{s}\}} \left\{\mathds{P}\left\{ \frac{w_\mathrm{s}}{N_\mathrm{s}+1} \log_2(1+\gamma_\mathrm{s})<U_\mathrm{s} \Big| A_\mathrm{s}, N_\mathrm{s}, d_\mathrm{s}, \right\}\right\} \\
            & \!=\! \int\limits_{0}\limits^{\infty} \int\limits_{0}\limits^{\infty} \sum\limits_{n=0}\limits^{\infty} \mathds{P} \left\{ \gamma_\mathrm{s}\! < \!2^{\frac{(n+1)U_\mathrm{s}}{w_\mathrm{s}}}\!-\!1 \right\} p_{N_\mathrm{s}}(n) f_{A_\mathrm{s}}(a) f_{d_\mathrm{s}}(d) \mathrm{d} a \mathrm{d} d,
        \end{split}
    \end{equation}
    where $p_{N_\mathrm{s}}(n)$, $f_{A_\mathrm{s}}(a)$ and $f_{d_\mathrm{s}}(d)$ denote the probability distribution functions of $N_\mathrm{s}$, $A_\mathrm{s}$ and $d_\mathrm{s}$ respectively.

    \vspace{2mm}
    \emph{\textbf{(1) No SC Sleeping}}

    Firstly, we analyze the outage probability when all SCs are active.
    In this case, $N_\mathrm{s}$ follows Poisson distribution of $A_\mathrm{s} \lambda_\mathrm{s}$, where $\lambda_\mathrm{s}$ is the density of SC UEs.
    Besides, $A_\mathrm{s}$ follows Gamma distribution with shape $K=3.575$ and scale $\frac{1}{K \rho_\mathrm{s}}$, where $\rho_\mathrm{s}$ is the density of SCs \cite{Slivnyak_theorem}.

    Notice that SCs will be more densely deployed in the future to boost network capacity, and the inter-cell interference, instead of noise, will be the main factor influencing the received SINR of SC UEs.
    In this case, we can derive the approximated outage probability as Theorem~2.

    \textbf{Theorem~2.} If $U_\mathrm{s}/w_\mathrm{s} \rightarrow 0$ and the SC-layer is interference-limited, the outage probability of the SC UEs without SC sleeping is given by:
        \begin{equation}\label{eq_outage_PBS_1}\footnotesize
            G_\mathrm{s} = 1- \frac{\frac{\alpha_\mathrm{s}-2}{2} 2^ {-\frac{U_\mathrm{s}}{w_\mathrm{s}} (1+\frac{\lambda_\mathrm{s}}{\rho_\mathrm{s}})}}{ 1 - \frac{4-\alpha_\mathrm{s}}{2} 2^{-\frac{U_\mathrm{s}}{w_\mathrm{s}}(1+\frac{\lambda_\mathrm{s}}{\rho_\mathrm{s}}) }  }.
        \end{equation}
    \emph{Proof}: See Appendix~\ref{appendix_TBS}.
    \hfill \rule{4pt}{8pt}\\

    Similarly, $U_\mathrm{s}/w_\mathrm{s} \rightarrow 0$ means the bandwidth is relatively large compared with the data rate requirement.
    Based on Theorem~2, the outage constraint for SC UEs without SC sleeping is equivalent to
        \begin{equation}\label{eq_outage_PBS_no_sleep_simple}\footnotesize
            \bar{w}_\mathrm{s} \log_2 \left( 1 + \tau_\mathrm{s} \right) \geq U_\mathrm{s},
        \end{equation}
    where $\bar{w}_\mathrm{s}=\frac{w_\mathrm{s}}{1+\bar{N}_\mathrm{s}}$ is the average bandwidth allocated to each SC UE, and $\tau_\mathrm{s}$ is defined as
        \begin{equation}\label{eq_tau_PBS_no_sleep}\footnotesize
            \tau_\mathrm{s} = \frac{\alpha_\mathrm{s}-2}{2} \frac{\eta_\mathrm{s}}{1-\eta_\mathrm{s}},
        \end{equation}
    denoting the received SINR of cell edge UEs of the SC-layer.

    \vspace{2mm}
    \emph{\textbf{(2) Repulsive Scheme}}

    Under the repulsive scheme, the distributions of $N_\mathrm{s}$, $A_\mathrm{s}$ and $d_\mathrm{s}$ remain the same after some SCs are turned off.
    As for the received SINR, only the UEs which locate around the sleeping area enjoy reduced inter-cell interference, while the received SINR of the other UEs is barely influenced.
    Therefore, we use Eq.~(\ref{eq_outage_PBS_no_sleep_simple}) to approximate the outage constraint of SC UEs by ignoring the benefit brought by SC sleeping.
    This is a conservative approximation of the real case.

    \vspace{2mm}
    \emph{\textbf{(3) Random Scheme}}

    Similarly to the repulsive scheme, the distributions of $N_\mathrm{s}$, $A_\mathrm{s}$ and $d_\mathrm{s}$ are not influenced by SC sleeping under the random scheme.
    However, the inter-cell interference received by a typical SC UE decreases by $p_\mathrm{s}$ on average.
    If the network is still interference-limited after SC sleeping, then noise can be ignored and the received SINR of the SC UEs will increase by $1/(1-p_\mathrm{s})-1$.
    Nonetheless, the received SINR will finally level off as the network becomes noise-limited.

    Inspired by Theorem~2, we use the following inequality to approximate the outage constraint of SC UEs under the random scheme:
        \begin{equation}\label{eq_outage_PBS_random_simple}\footnotesize
            \bar{w}_\mathrm{s} \log_2 \left( 1 + \tau'_\mathrm{s}(p_\mathrm{s}) \right) \geq U_\mathrm{s},
        \end{equation}
    where $\tau'_\mathrm{s}(p_\mathrm{s})$ is defined as
    \begin{equation}\label{eq_tau_PBS}\footnotesize
        \tau'_\mathrm{s}(p_\mathrm{s}) = \frac{\alpha_\mathrm{s}-2}{2(1-\min(\hat{p}_\mathrm{s}, p_\mathrm{s}))} \frac{\eta_\mathrm{s}}{1-\eta_\mathrm{s}},
    \end{equation}
    and $\hat{p}_\mathrm{s}$ is an experimental threshold of $p_\mathrm{s}$, indicating whether the network is interference-limited or not.
    Specifically, the network is considered as noise-limited if $\hat{p}_\mathrm{s}<p_\mathrm{s}$, in which case turning off SCs no longer improve the received SINR.
    In fact, the additional noise is approximated by the inter-cell interference of $p_\mathrm{s}=\hat{p}_\mathrm{s}$ in Eq.~(\ref{eq_tau_PBS}).
    Thus the approximated QoS constraint Eq.~(\ref{eq_outage_PBS_random_simple}) is more strict for smaller $\hat{p}_\mathrm{s}$.
    Notice that $\tau'_\mathrm{s}(p_\mathrm{s})$ reflects both inter-cell interference and noise.

    \subsection{Outage Constraint of Offloaded UEs}

    \emph{\textbf{(1) No CB}}

    Under the random scheme, the outage probability of the offloaded UEs can be obtained in the same way as Theorem~1, and the outage QoS is equivalent to:
        \begin{equation}\label{eq_outage_offloaded_random_no_CA}\footnotesize
            \frac{W_\mathrm{m}-w_\mathrm{m}}{1+ \frac{3\sqrt{3}}{2}\lambda_\mathrm{s} p_\mathrm{s} D^2} \log_2 \left( 1 + \frac{P^\mathrm{m}}{\sigma^2(1+I)} \frac{\alpha_\mathrm{m}+2}{2} \frac{\eta_\mathrm{o}}{D^{\alpha_\mathrm{m}}} \right) \geq U_\mathrm{o}.
        \end{equation}
    Similarly, the outage constraint of the offloaded UEs under repulsive scheme can also be simplified:
         \begin{equation}\label{eq_outage_offloaded_repulsive_no_CA}\footnotesize
            \frac{W_\mathrm{m}-w_\mathrm{m}}{1+ \pi R_\mathrm{s}^2 \lambda_\mathrm{s}} \log_2 \left( 1 + \frac{P^\mathrm{m}}{\sigma^2(1+I)} \frac{\alpha_\mathrm{m}+2}{2} \frac{\eta_\mathrm{o}}{R_\mathrm{s}^{\alpha_\mathrm{m}}} \right) \geq U_\mathrm{o}.
        \end{equation}

    \vspace{2mm}
    \emph{\textbf{(2) With CB}}

    In this case, there are two bands serving the offloaded UEs which may have different path loss factors: $W_\mathrm{m}-w_\mathrm{m}$ and $W_\mathrm{s}-w_\mathrm{s}$.
    For simplicity, we assume the offloaded UEs are divided into two groups randomly, and each group shares one band.

    For the random scheme, the outage constraint for the offloaded UEs is as follows:
        \begin{equation}\label{eq_outage_offloaded_random_CA} \footnotesize
        \left\{
        \begin{split}
            & \frac{W_\mathrm{m}-w_\mathrm{m}}{1+ \frac{3\sqrt{3}}{2}\lambda_\mathrm{s} p_\mathrm{s} p_\mathrm{m} D^2} \log_2 \left( 1 + \tau_\mathrm{o}(\alpha_\mathrm{m}, D) \right) \geq U_\mathrm{o}\\
            & \frac{W_\mathrm{s}-w_\mathrm{s}}{1+ \frac{3\sqrt{3}}{2}\lambda_\mathrm{s} p_\mathrm{s} (1-p_\mathrm{m}) D^2} \log_2 \left( 1 + \tau_\mathrm{o}(\alpha_\mathrm{s}, D) \right) \geq U_\mathrm{o},
        \end{split}
        \right.
        \end{equation}
    where $p_\mathrm{m}$ is the probability that an offloaded UE uses the bandwidth of the MBS-layer, $\tau_\mathrm{o}(\alpha, r)$ is defined as
        \begin{equation}\label{eq_tau_o}\footnotesize
            \tau_\mathrm{o}(\alpha,r)=\frac{P^\mathrm{m}}{\sigma^2(1+I)} \frac{\alpha+2}{2} \frac{\eta_\mathrm{o}}{r^{\alpha}}.
        \end{equation}
    $\tau_\mathrm{o}(\alpha, r)$ is a threshold of the received SINR of the offloaded UEs, when they are uniformly distributed within circles of radius $r$ centered at MBSs with the path loss factor $\alpha$.

    Similarly, the outage constraint of the offloaded UEs under the repulsive scheme is given by:
        \begin{equation}\label{eq_outage_offloaded_repulsive_CA}\footnotesize
        \left\{
        \begin{split}
            & \frac{W_\mathrm{m}-w_\mathrm{m}}{1+ \pi R_\mathrm{s}^2 \lambda_\mathrm{s} p_\mathrm{m}} \log_2 \left( 1 + \tau_\mathrm{o}(\alpha_\mathrm{m},R_\mathrm{s}) \right) \geq U_\mathrm{o} \\
            & \frac{W_\mathrm{s}-w_\mathrm{s}}{1+ \pi R_\mathrm{s}^2 \lambda_\mathrm{s} (1-p_\mathrm{m})} \log_2 \left( 1 + \tau_\mathrm{o}(\alpha_\mathrm{s},R_\mathrm{s} \right) \geq U_\mathrm{o}.
        \end{split}
        \right.
        \end{equation}

    To achieve load balance of the two bands and turn off more SCs, the probability $p_\mathrm{m}$ is also considered as a variable to be optimized.

\subsection{Evaluations of Analytical Outage Probabilities}
    \begin{table}[!t]
        \caption{Simulation Parameters}
        \label{tab_parameter}
        \centering
        \begin{tabular}{||c|c||c|c||}
        \hline
        Parameter & Value & Parameter & Value\\
        \hline
        $D$ & 500m & $\rho_\mathrm{s}$ & 25/km$^2$\\
        $P^\mathrm{m}$ & 10W & $P^\mathrm{s}$ & 1W \\
        $W_\mathrm{m}$ & 10MHz & $W_\mathrm{s}$ & 10MHz \\
        $\sigma^2$ & -104dBm & $\alpha_\mathrm{s}$ & 4\\
        $U_\mathrm{m}$ & 64kbps & $\eta_\mathrm{m}$ & 0.05 \\
        $U_\mathrm{s}$ & 100kbps & $\eta_\mathrm{s}$ & 0.05 \\
        $U_\mathrm{o}$ & 100kbps & $\eta_\mathrm{o}$ & 0.05 \\
        % $P_0^\mathrm{m}$ & 56W & $P_0^\mathrm{s}$ & 10W \\
        % $\delta_\mathrm{m}$ & 4 & $\delta_\mathrm{s}$ & 4 \\
        \hline
        \end{tabular}
    \end{table}

Now we evaluate the approximation errors of the derived closed-form outage probabilities under the two schemes.
A HCN with 19 MBSs is considered for simulation, where the number of SCs (and UEs) and their locations are set up by Monte Carlo simulation method.
The simulation parameters are listed in Table~\ref{tab_parameter} \cite{WZhang_separation}.
By calculating the data rate of each UE, the outage probability can be obtained.

When $w_\mathrm{m}=w_\mathrm{s}=w_\mathrm{o}=10$MHz, the simulation and analytical results are compared in Fig.~\ref{fig_evaluation}.
The analytical results of the four sub-figures come from Eqs.~(\ref{eq_outage_MBS_simple}), (\ref{eq_outage_PBS_random_simple}), (\ref{eq_outage_offloaded_repulsive_no_CA}), and (\ref{eq_outage_PBS_no_sleep_simple}) respectively.
Note that the analytical results are quite close to the simulation results in Fig.~\ref{fig_evaluation}(a-c), which validates the corresponding assumptions and approximations.
However, the error of the analytical results in Fig.~\ref{fig_evaluation}(d) increases with the sleeping radius.
This is because of the conservative approximation that uses Eq.~(\ref{eq_outage_PBS_no_sleep_simple}) to calculate the outage probability of SC UEs under the repulsive scheme, which exaggerates the inter-cell interference.
Nevertheless, the analytical outage probability is a upper bound of the real case, based on which the sub-optimal solution can be derived.

%% file: ProblemSolution.tex
Based on the derived outage constraints in the last section, we analyze how many SCs can be turned off under the random and repulsive schemes respectively.

\subsection{Optimal Random Scheme}
    Under the random scheme, the problem is formulated as follows.
    \begin{eqnarray}\label{eq_formulation_random} \footnotesize
        \max\limits_{p_\mathrm{s},p_\mathrm{m}} & & p_\mathrm{s} \nonumber\\
        \mbox{s.t.} & &  \frac{w_\mathrm{s}}{1\!+\!\frac{\lambda_\mathrm{s}}{\rho_\mathrm{s}}} \log_2 \left( 1\! +\! \tau'(p_\mathrm{s}) \right) \geq U_\mathrm{s} \nonumber\\
        & & \frac{w_\mathrm{m}}{1\!+\! \frac{3\sqrt{3}}{2}\lambda_\mathrm{m} D^2} \log_2 \left( 1 \!+\! \tau_\mathrm{m} \right) \geq U_\mathrm{m} \nonumber \\
        & & \frac{W_\mathrm{m}-w_\mathrm{m}}{1\!+\! \frac{3\sqrt{3}}{2}\lambda_\mathrm{s} p_\mathrm{s} p_\mathrm{m} D^2} \log_2 \left( 1 \!+\! \tau_\mathrm{o}(\alpha_\mathrm{m},D) \right) \geq U_\mathrm{o}\\
        & & \frac{W_\mathrm{s}-w_\mathrm{s}}{1\!+\! \frac{3\sqrt{3}}{2}\lambda_\mathrm{s} p_\mathrm{s} (1-p_\mathrm{m}) D^2} \log_2 \left( 1\! +\! \tau_\mathrm{o}(\alpha_\mathrm{s},D)  \right) \geq U_\mathrm{o} \nonumber \\
        & & p_\mathrm{m} \left\{ \begin{array}{ll} \in (0,1), & \mbox{with CB} \\ = 1, & \mbox{without CB} \end{array} \right. \nonumber,
    \end{eqnarray}
    where $\tau_\mathrm{m}$, $\tau'_\mathrm{s}(p_\mathrm{s})$, and $\tau_\mathrm{o}(\alpha,r)$ are given by Eqs.~(\ref{eq_tau_MBS}), (\ref{eq_tau_PBS}), (\ref{eq_tau_o}) respectively.
    The equality of the service outage constraints should hold under the optimal solution, and thus all the residual bandwidth can be utilized to turn off more SCs.

    \vspace{2mm}
    \emph{\textbf{(1) No CB}}

    When CB is not conducted at MBSs, $p_\mathrm{m}=1$.
    In this case, the first and fourth conditions are invalid, and the optimal value of this problem is given by
        \begin{equation}\label{eq_q_opt_rand_no_CA} \footnotesize
        \begin{split}
            p_\mathrm{s}^* & = \frac{\rho_\mathrm{m}}{\lambda_\mathrm{s}} \left( \frac{\log_2(1 + \tau_\mathrm{o}(\alpha_\mathrm{m},D))}{U_\mathrm{o}} \left( W_\mathrm{m} - \frac{U_\mathrm{m}(1+\frac{\lambda_\mathrm{m}}{\rho_\mathrm{m}})}{\log_2(1 + \tau_\mathrm{m})} \right)  -1 \right),
            %\left( \frac{\log_2(1+\tau_\mathrm{m})}{U_\mathrm{o}} W_\mathrm{m} - \frac{U_\mathrm{m} \log_2(1 + \tau_\mathrm{o}(\alpha_\mathrm{m},D))}{U_\mathrm{o} \log_2(1 + \tau_\mathrm{m})} - 1 \right) \frac{1}{\lambda_\mathrm{s}} \rho_\mathrm{m} - \frac{\lambda_\mathrm{m}}{\lambda_\mathrm{s}} \frac{U_\mathrm{m}\log_2(1 + \tau_\mathrm{o}(\alpha_\mathrm{m},D))}{U_\mathrm{o}\log_2(1 + \tau_\mathrm{m})},
        \end{split}
        \end{equation}
    where $\rho_\mathrm{m}=\frac{1}{\frac{3\sqrt{3}}{2}D^2}$.\\

    \emph{\textbf{Proposition~1}} The ratio of sleeping SCs is inversely proportional to the density of SC UEs $\lambda_\mathrm{s}$, and decreases linearly with the density of MBS UEs $\lambda_\mathrm{m}$.

    \emph{\textbf{Proposition~2}} Deploying denser MBSs can help to save energy, whereas the density of SCs should be minimized, i.e., just satisfying the peak traffic demand of high data rate service.\\
    
    \begin{figure}[!t]
        \centering
        \subfloat[MBS UEs] {\includegraphics[width=1.6in]{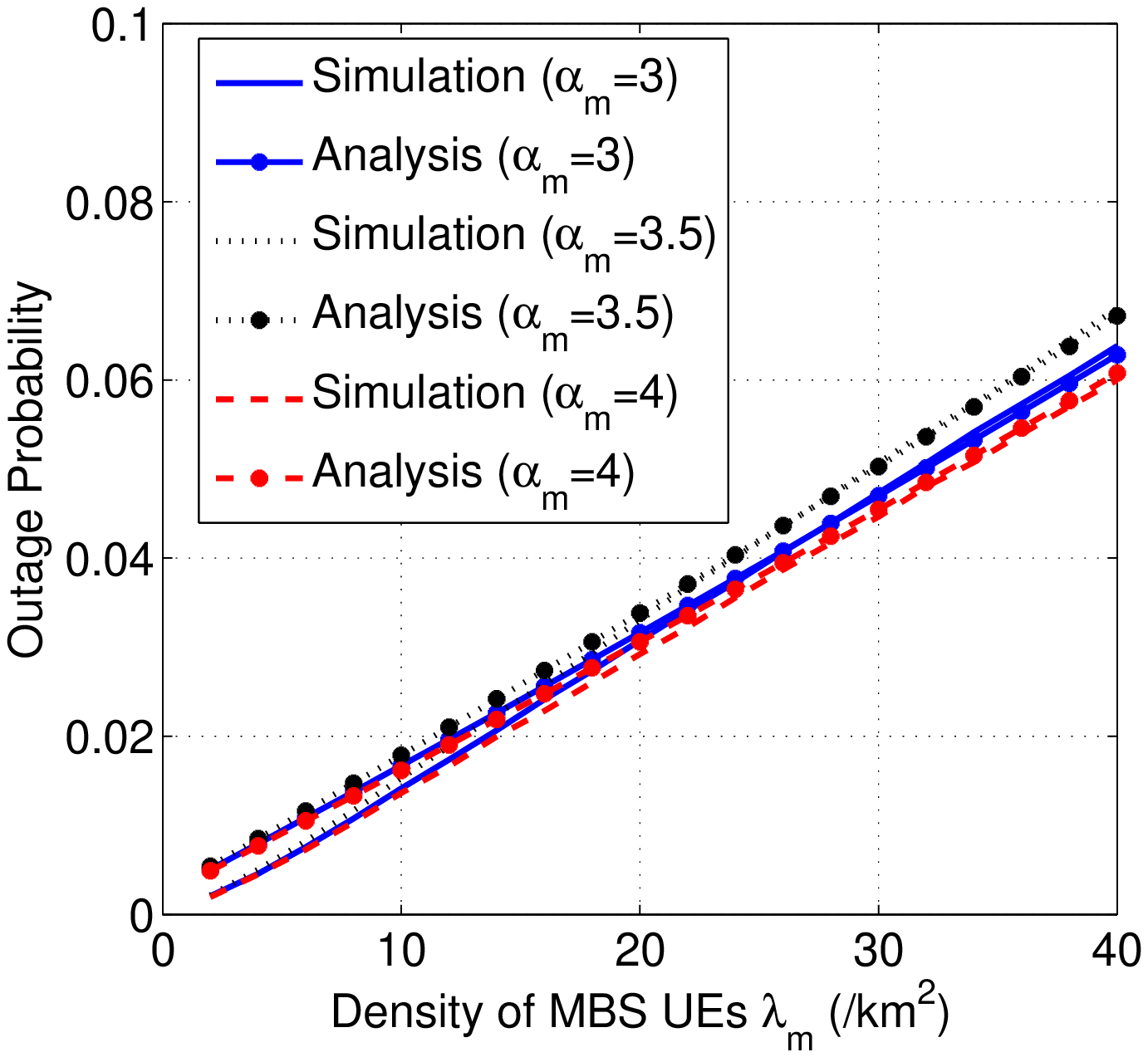}
        \label{fig_evaluation_MBS}}
        %\hfil
        \subfloat[SC UEs (random scheme)]{\includegraphics[width=1.6in]{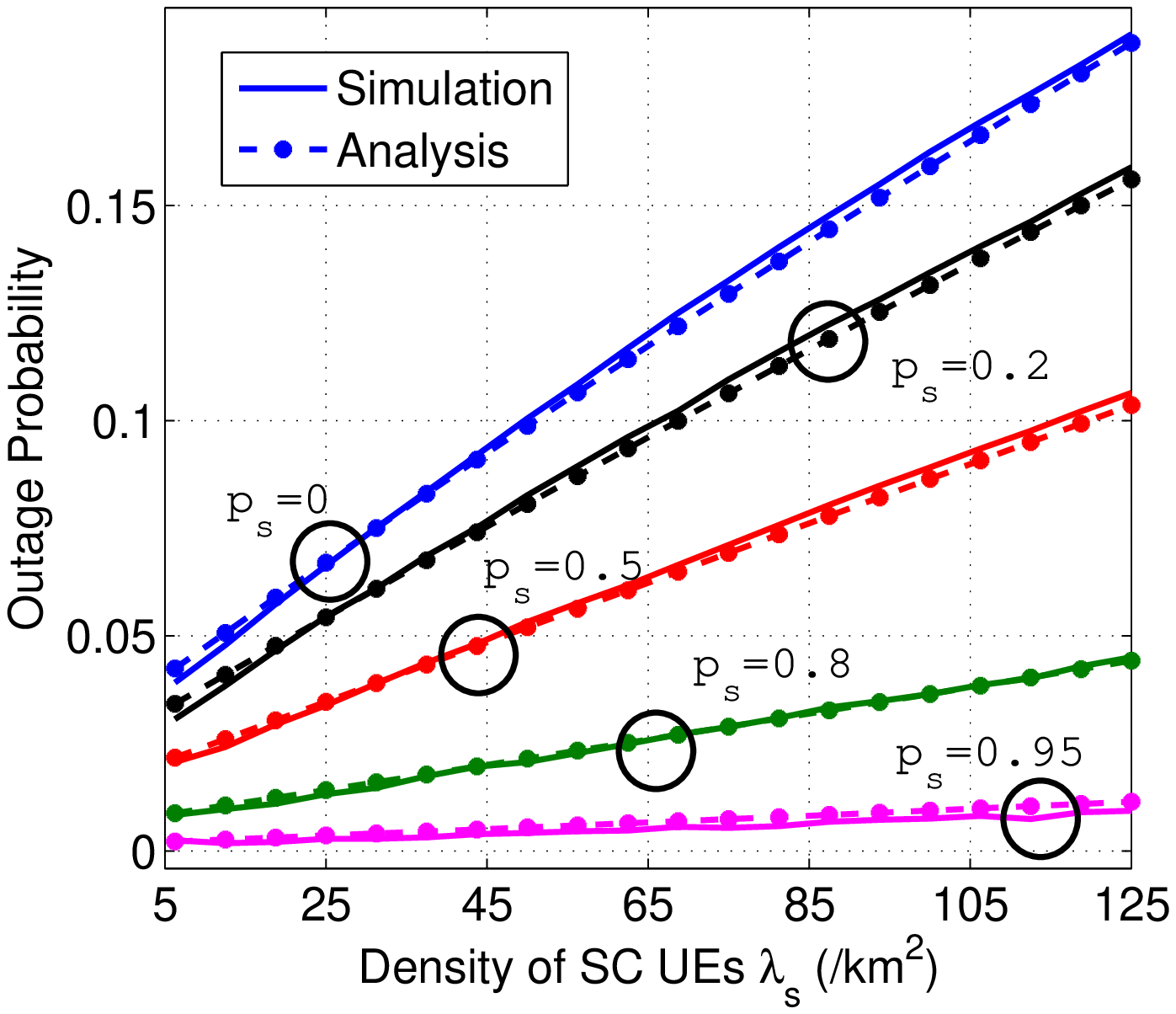}
        \label{fig_evaluation_PBS_rand}}
        \hfil\\
        \subfloat[offloaded UEs (repulsive scheme: $R_\mathrm{s}=300$m)] {\includegraphics[width=1.6in]{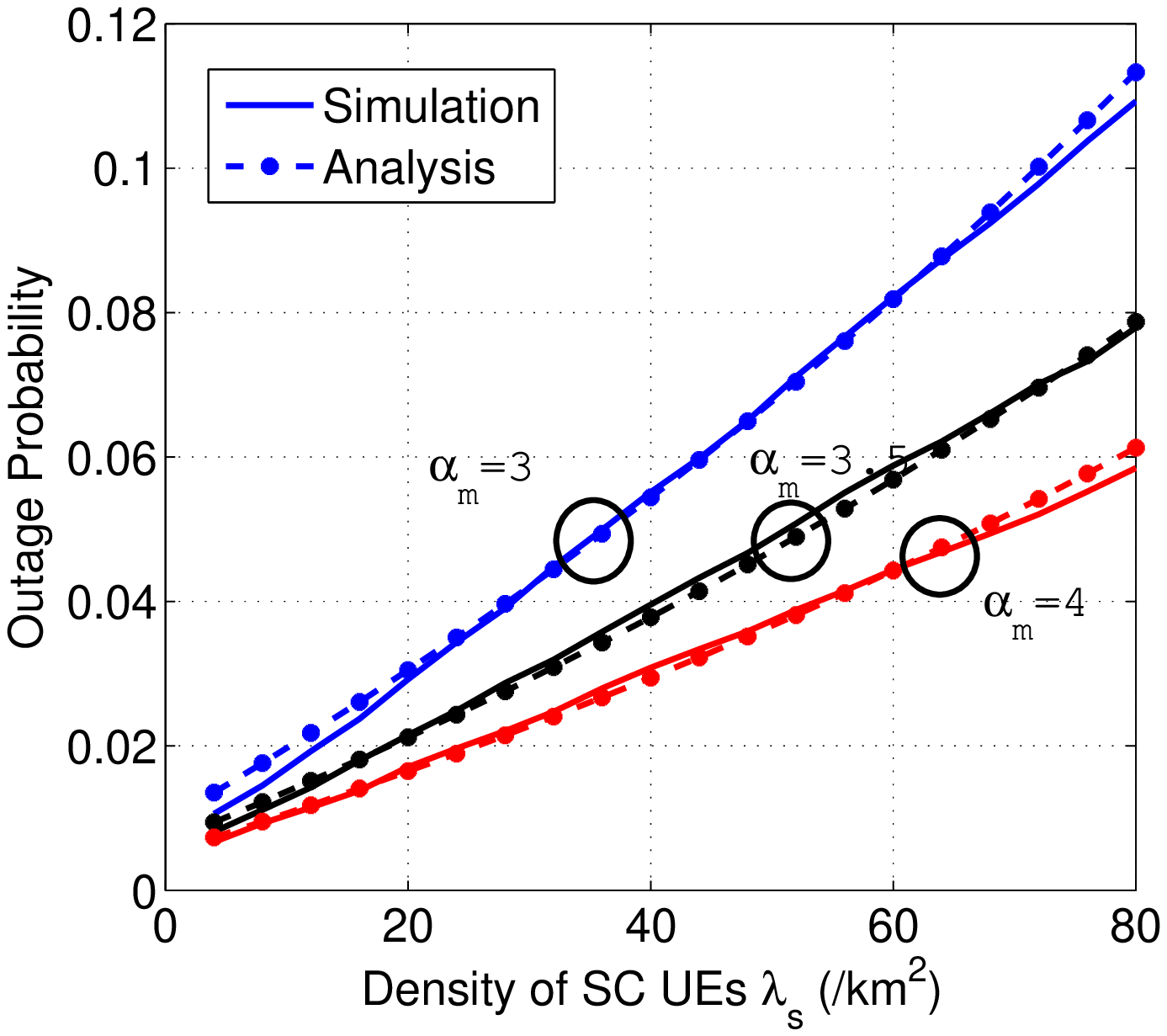}
        \label{fig_evaluation_offloaded_repulsive}}
        %\hfil
        \subfloat[SC UEs (repulsive scheme)]{\includegraphics[width=1.6in]{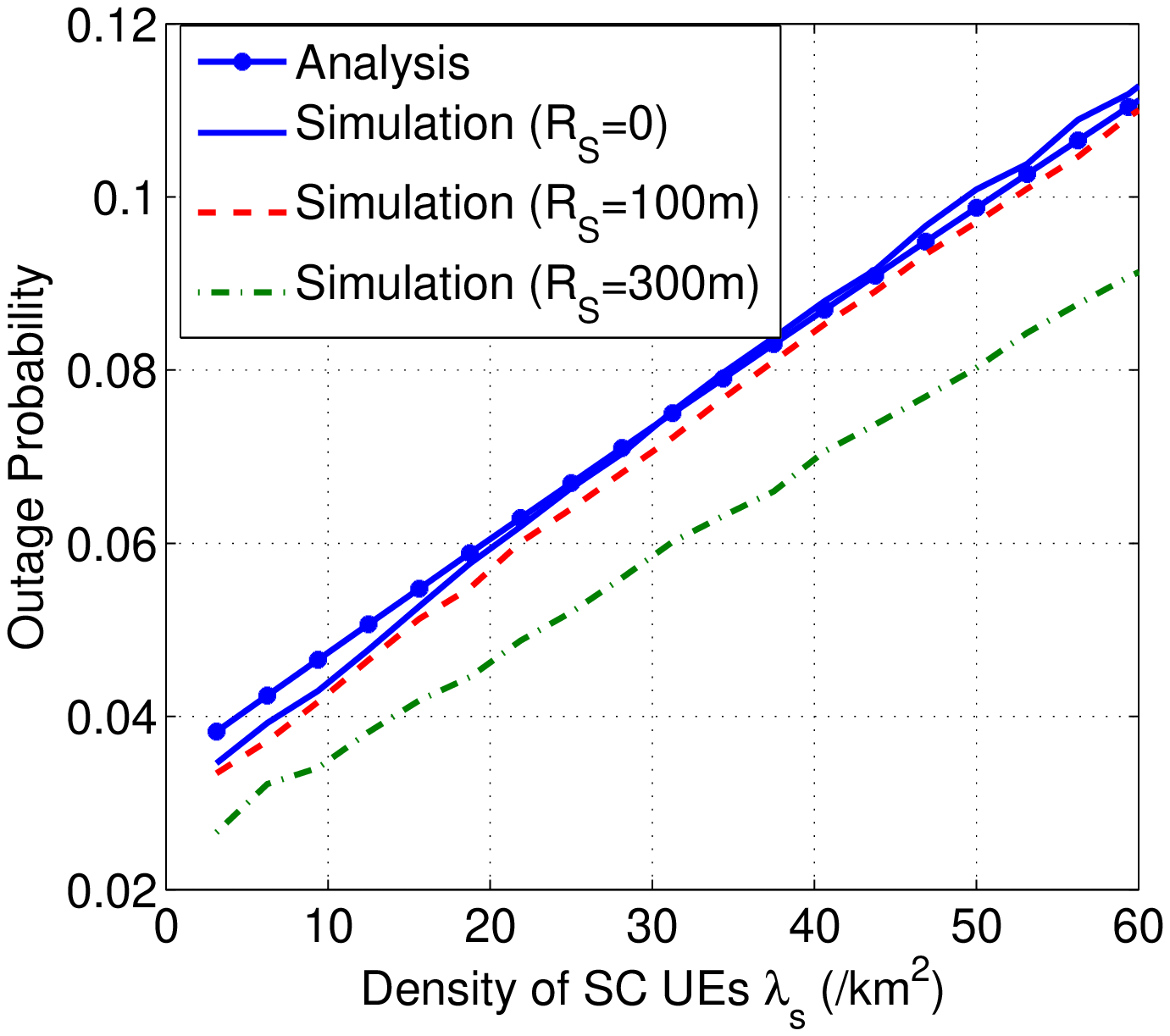}
        \label{fig_evaluation_PBS_repulsive}}
        \caption{Comparison of analytical results and simulation results.}
        \label{fig_evaluation}
    \end{figure}

    Notice that Proposition~1 roughly describes the amount of energy that can be saved due to the dynamics of traffic load in time domain.
    Recall that the bandwidth for the offloaded UEs decreases linearly with the density of MBS UEs based on Eq.~(\ref{eq_outage_MBS_simple}), therefore, the capability of the MBS-layer to handle the offloaded UEs also decreases linearly with $\lambda_\mathrm{m}$.
    Furthermore, the density of the offloaded UE given by $p_\mathrm{s} \lambda_\mathrm{s}$ is limited by the offloading capability of the MBS-layer, thus the sleeping ratio is inversely proportional to the traffic load of the SC-layer.

    Proposition~2 offers insights for energy-efficient network deployment.
    According to the traditional network planning method, the density of MBSs and SCs should both be minimized to meet the peak hour traffic demands.
    However, this conclusion is inaccurate if SC sleeping and vertical offloading are introduced.
    The density of active SCs given by $(1-p_\mathrm{s})^* \rho_\mathrm{s}$ increases linearly with $\rho_\mathrm{s}$ as $p_\mathrm{s}^*$ is irrelevant with $\rho_\mathrm{s}$, which suggests that deploying more SCs only results in denser active SCs consuming more energy.
    Therefore, the traditional method is still energy-optimal for SC deployment.
    Whereas, deploying more MBSs helps to turn off more SCs and reduces the energy consumption of the SC-layer.
    Therefore, the energy-optimal density of MBSs may be larger than the one obtained by traditional method.

    \vspace{2mm}
    \emph{\textbf{(2) With CB}}

    When CB is conducted, the optimal solution has no closed-form expression as the first, third and fourth constraints are coupled together by $p_\mathrm{s}$.
    Only numerical results can be obtained by methods like dichotomy.

    To offer some insights, we consider a special case when $\alpha_\mathrm{m} = \alpha_\mathrm{s}$ (the same path loss factor of the two layers) and $p_\mathrm{s}\geq\hat{p}_\mathrm{s}$ (noise-limited case when many SCs are turned off).
    Under this condition, the three constraints can be decoupled and the optimal solution is
    \begin{equation} \footnotesize
        \label{eq_q_opt_rand_CA}
        \begin{split}
            {p'_\mathrm{s}}^* = & \frac{\rho_\mathrm{m}}{\lambda_\mathrm{s}} \left( \frac{\log_2(1 + \tau_\mathrm{o}(\alpha_\mathrm{m},D))}{U_\mathrm{o}} \left( W_\mathrm{m}  - \frac{U_\mathrm{m}(1+\frac{\lambda_\mathrm{m}}{\rho_\mathrm{m}})}{\log_2(1 + \tau_\mathrm{m})} \right. \right.\\
            & \left. \left. + W_\mathrm{s} - \frac{U_\mathrm{s}(1+\frac{\lambda_\mathrm{s}}{\rho_\mathrm{s}})}{\log_2(1 + \tau'_\mathrm{s}(\hat{p}_\mathrm{s}))} \right) - 1 \right).
        \end{split}
    \end{equation}
    Then, the performance gain by introducing CB is given by
    \begin{equation} \footnotesize
        \label{eq_q_differnce_rand_CA}
            {p'_\mathrm{s}}^* \!-\! {p_\mathrm{s}}^* \!=\! \frac{\rho_\mathrm{m}}{\lambda_\mathrm{s}} \left( \frac{\log_2(1 \!+\! \tau_\mathrm{o}(\alpha_\mathrm{m},D))}{U_\mathrm{o}} \left( W_\mathrm{s} \!-\! \frac{U_\mathrm{s}(1+\frac{\lambda_\mathrm{s}}{\rho_\mathrm{s}})}{\log_2(1 \!+\! \tau'_\mathrm{s}(\hat{p}_\mathrm{s}))} \right) \right),
    \end{equation}
    which is inversely proportional to the density of SC UEs and increases linearly with the redundant bandwidth at SC-layer.\\

    \emph{\textbf{Proposition~3}} CB is more beneficial when the traffic load of the SC-layer is lower.

    \emph{\textbf{Proposition~4}} Denser networks (with larger $\rho_\mathrm{m}$ and $\rho_\mathrm{s}$) will benefit more from conducting CB.\\

    Denser SCs and lower traffic load of SC-layer both means more redundant bandwidth at SC-layer, and thus conducting CB will bring higher capacity gains for the offloaded UEs.
    Similarly, networks with denser MBSs can make more use of the borrowed bandwidth to increase the network capacity.

\subsection{Optimal Repulsive Scheme}

    The problem under the repulsive scheme can be formulated as follows.
    \begin{eqnarray}\label{eq_formulation_repulsive} \footnotesize
        \max\limits_{R_\mathrm{s},p_\mathrm{m}} & & \pi R_\mathrm{s}^2 \rho_\mathrm{m} \nonumber\\
        \mbox{s.t.} & &  \frac{w_\mathrm{s}}{1\!+\!\frac{\lambda_\mathrm{s}}{\rho_\mathrm{s}}} \log_2 \left( 1\! + \!\tau_\mathrm{s}\right) \geq U_\mathrm{s} \nonumber\\
        & & \frac{w_\mathrm{m}}{1\!+\! \frac{3\sqrt{3}}{2}\lambda_\mathrm{m} D^2} \log_2 \left( 1 \!+\! \tau_\mathrm{m} \right) \geq U_\mathrm{m} \nonumber \\
        & & \frac{W_\mathrm{m}-w_\mathrm{m}}{1\!+\! \pi R_\mathrm{s}^2 \lambda_\mathrm{s} p_\mathrm{m}} \log_2 \left( 1 \!+\! \tau_\mathrm{o}(\alpha_\mathrm{m},R_\mathrm{s}) \right) \geq U_\mathrm{o} \\
        & & \frac{W_\mathrm{s}-w_\mathrm{s}}{1 \!+\! \pi R_\mathrm{s}^2 \lambda_\mathrm{s} (1-p_\mathrm{m})} \log_2 \left( 1 \!+\! \tau_\mathrm{o}(\alpha_\mathrm{s},R_\mathrm{s}) \right) \geq U_\mathrm{o} \nonumber\\
        & & p_\mathrm{m} \left\{ \begin{array}{ll} \in (0,1), & \mbox{with CB} \\ = 1, & \mbox{without CB} \end{array} \right. \nonumber.
    \end{eqnarray}
    The four service outage constraints should take equality under the optimal solution, and we have
    \begin{equation}\label{eq_band_opt_repulsive}\footnotesize
            w_\mathrm{s} = \frac{U_\mathrm{s} (1 + \frac{\lambda_\mathrm{s}}{\rho_\mathrm{s}})}{\log_2(1+\tau_\mathrm{s})},
            w_\mathrm{m} = \frac{U_\mathrm{m} (1 + \frac{\lambda_\mathrm{m}}{\rho_\mathrm{m}})}{\log_2(1+\tau_\mathrm{m})}.
    \end{equation}

    \vspace{2mm}
    \emph{\textbf{(1) Without CB}}

    When CB is not conducted, the optimal solution $R_\mathrm{s}^*$ should satisfy
    \begin{equation}\label{eq_con_repulsive_no}\footnotesize
        \frac{W_\mathrm{m}-\frac{U_\mathrm{m} (1 + \frac{\lambda_\mathrm{m}}{\rho_\mathrm{m}})}{\log_2(1+\tau_\mathrm{m})}}{1\!+\! \pi {R_\mathrm{s}^*}^2 \lambda_\mathrm{s} p_\mathrm{m}} \log_2 \left( 1 \!+\! \tau_\mathrm{o}(\alpha_\mathrm{m},{R_\mathrm{s}^*}^2) \right) = U_\mathrm{o},
    \end{equation}
    based on which the numerical results of Eq.~(\ref{eq_formulation_repulsive}) can be obtained.
    Note that Proposition~2 still holds as $R_\mathrm{s}^*$ is irrelevant with $\rho_\mathrm{s}$ whereas increases with $\rho_\mathrm{m}$.\\

    The upper bound of the optimal value of problem Eq.~(\ref{eq_formulation_repulsive}) can be derived by rewriting the constraint of the offloaded UEs as
    \begin{equation}\label{eq_upper_bound_constraint} \footnotesize
    \begin{split}
        & \frac{P^\mathrm{m}}{\sigma^2(1+I)} \frac{\alpha_\mathrm{m}+2}{2} \eta_\mathrm{c} \geq \left(2^{\frac{(1+\pi R_\mathrm{s}^2 \lambda_\mathrm{s})U_\mathrm{o}}{ W_\mathrm{m}\!-\!w_\mathrm{m}}} - 1\right) R_\mathrm{s}^\mathrm{\alpha_\mathrm{m}} \\
        \overset{\mbox{(a)}}{\geq} &\frac{U_\mathrm{o} \ln2} {W_\mathrm{m}\!-\!w_\mathrm{m}} (1+\pi R_\mathrm{s}^2 \lambda_\mathrm{s}) R_\mathrm{s}^\mathrm{\alpha_\mathrm{m}} > \frac{U_\mathrm{o} \ln2} {W_\mathrm{m}-w_\mathrm{m}} \pi \lambda_\mathrm{s} R_\mathrm{s}^{\mathrm{\alpha_\mathrm{m}}+2},
    \end{split}
    \end{equation}
    where (a) applies the inequality $e^{x}-1\geq x(x\geq0)$.
    Then the upper bound of SC sleeping ratio is $\pi \tilde{R}_\mathrm{s}^2 \rho_\mathrm{m}$, with the maximum sleeping radius $\tilde{R}_\mathrm{s}$ given by
    \begin{equation}\label{eq_opt_r_no_CA} \footnotesize
        \tilde{R}_\mathrm{s} = \left( \frac{(\alpha_\mathrm{m}+2)\eta_\mathrm{o}P^\mathrm{m}}{2\sigma^2(1+I)U_\mathrm{o} \pi \lambda_\mathrm{s}} \left(W_\mathrm{m} - \frac{U_\mathrm{m} (1 + \frac{\lambda_\mathrm{m}}{\rho_\mathrm{m}})}{\log_2(1+\tau_\mathrm{m})}\right) \right)^{\frac{1}{\alpha_\mathrm{m}+2}}.
    \end{equation}
    Specially, this bound is quite tight as $U_\mathrm{o}/(W_\mathrm{m}-w_\mathrm{m}) \rightarrow 0$.\\

    \emph{\textbf{Proposition~5}} The repulsive scheme is more benificial for the heavily loaded networks.\\

    Eq.~(\ref{eq_opt_r_no_CA}) indicates that the ratio of sleeping SCs under the repulsive scheme is inversely proportional to $\rho_\mathrm{s}^{\frac{2}{\alpha_\mathrm{m}+2}}$.
    As $\alpha_\mathrm{m} \in (2,4]$, the performance of the repulsive scheme is less sensitive to the variation of the traffic load compared with the random scheme, which explains Proposition~5.
    In fact, the advantages of the repulsive scheme mainly comes from the shorter distance and smaller path loss of the offloaded UEs.
    However, this advantage degrades as $R_\mathrm{s}$ increases.
    That is why the repulsive scheme is more advantageous for the heavily loaded networks where few SCs can be turned off and the sleeping radius $R_\mathrm{s}$ is small.

    \vspace{2mm}
    \emph{\textbf{(2) With CB}}

    If CB is supported and $\alpha_\mathrm{m} = \alpha_\mathrm{s}$, the available bandwidth for the offloaded UEs $w_\mathrm{o} = W_\mathrm{m} \!+\! W_\mathrm{s} \!-\! w_\mathrm{m} - w_\mathrm{s}$, where $w_\mathrm{m}$ and $w_\mathrm{s}$ are given by Eq.~(\ref{eq_band_opt_repulsive}).
    In this case, the upper bound of the sleeping radius becomes
    \begin{equation}\label{eq_opt_r_CA}\footnotesize
        \tilde{R'}_\mathrm{s} \!=\! \left( \frac{(\alpha_\mathrm{m}\!+\!2)\eta_\mathrm{o}P^\mathrm{m}}{2\sigma^2(1\!+\!I)U_\mathrm{o} \pi \lambda_\mathrm{s}} \left(W_\mathrm{m} \!+\! W_\mathrm{s} \!-\! w_\mathrm{m} - w_\mathrm{s}\right) \right)^{\frac{1}{\alpha_\mathrm{m}\!+\!2}}.
    \end{equation}

    Notice that Proposition~3 and Proposition~4 also hold for the repulsive scheme.

\subsection{Numerical Results and Analysis}
    \begin{table}[!t]
        \caption{Network density}
        \label{tab_density}
        \centering
        \begin{tabular}{||c||c|c|c||}
        \hline
        Parameter & Baseline & Denser SCs & Denser MBSs\\
        \hline
        $D$ & 500m & 500m & 400m \\
        $\rho_\mathrm{s}$ & 25/km$^2$ & 50/km$^2$ & 25/km$^2$ \\
        \hline
        \end{tabular}
    \end{table}

    \begin{figure}[!t]
        \centering
        \subfloat[Varying density of SC UEs $\lambda_\mathrm{s}$ ($\lambda_\mathrm{m}=20$/km$^2$)] {\includegraphics[width=2.5in]{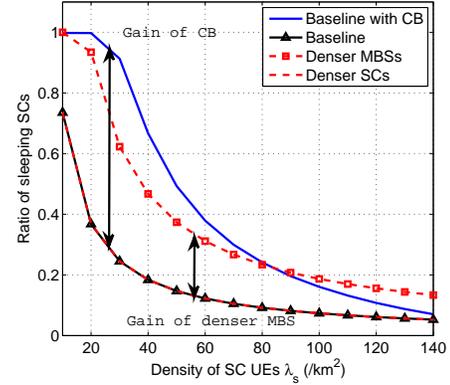}
        \label{fig_rand_lambda_p}}
        \hfil\\
        \subfloat[Varying density of MBS UEs $\lambda_\mathrm{m}$ ($\lambda_\mathrm{s}=100$/km$^2$)] {\includegraphics[width=2.5in]{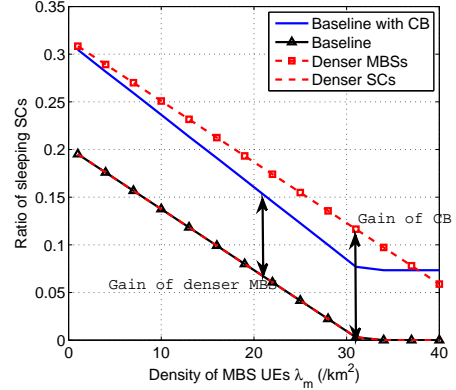}
        \label{fig_rand_lambda_m}}
        \caption{Maximum ratio of sleeping SC under the random scheme.}
        \label{fig_rand}
    \end{figure}

    \begin{figure}[!t]
        \centering
        \subfloat[Varying density of SC UEs $\lambda_\mathrm{s}$ ($\lambda_\mathrm{m}=20$/km$^2$)] {\includegraphics[width=2.5in]{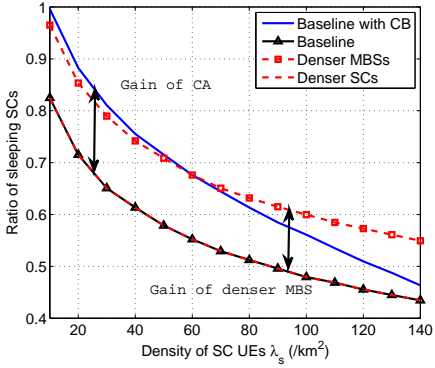}
        \label{fig_repulsive_lambda_p}}
        \hfil\\
        \subfloat[Varying density of MBS UEs $\lambda_\mathrm{m}$ ($\lambda_\mathrm{s}=100$/km$^2$)] {\includegraphics[width=2.5in]{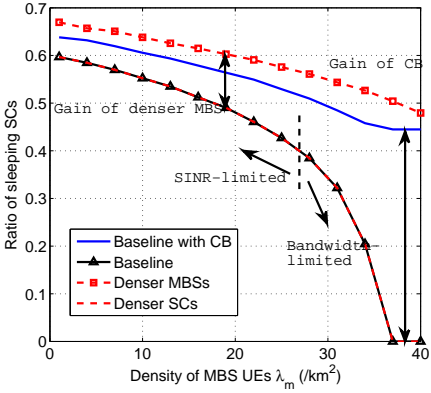}
        \label{fig_repulsive_lambda_m}}
        \caption{Maximum ratio of sleeping SC under the repulsive scheme.}
        \label{fig_repulsive}
    \end{figure}

    In this part, we analyze the relationship between the maximum sleeping ratio and system parameters (traffic load and BS density) through the optimal solutions of Problems~(\ref{eq_formulation_random}) and (\ref{eq_formulation_repulsive}) obtained by dichotomy.
    $\alpha_\mathrm{m}$ is set to 3.5, and other parameters can be found in Table~\ref{tab_parameter}.

    The maximum sleeping ratio versus traffic load $\lambda_\mathrm{s}$ and $\lambda_\mathrm{m}$ under random scheme is presented in Fig.~\ref{fig_rand}, where the influences of BS density and CB are both considered.
    The settings of network density is listed in Table~\ref{tab_density}.
    Besides, the results of the repulsive scheme are demonstrated in Fig.~\ref{fig_repulsive}
    Notice that the black solid lines with triangles are totally overlapped by the red dashed lines in Fig.~\ref{fig_rand} and Fig.~\ref{fig_repulsive}.

    \emph{\textbf{(1) Traffic load}}

    Generally, the sleeping ratio decreases with traffic load, whereas the slopes of the curves are quite different.
    Under the random scheme, the sleeping ratio is inversely proportional to the density of SC UEs in Fig.~\ref{fig_rand}(a), whereas it shows linear relation with the density of MBS UEs in Fig.~\ref{fig_rand}(b).
    Under the repulsive scheme, the sleeping ratio decreases more slowly with $\lambda_\mathrm{s}$ (Fig.\ref{fig_repulsive}(a)), but there is a rapid decline when the traffic load of the MBS is high (Fig.\ref{fig_repulsive}(b)).
    These results are consistent with the analytical results of Eqs.~(\ref{eq_q_opt_rand_no_CA}),(\ref{eq_q_opt_rand_CA}),(\ref{eq_opt_r_no_CA}), and (\ref{eq_opt_r_CA}).

    Notice that the sleeping ratio under the repulsive scheme can be divided into two cases:
    (1) high sleeping ratio with large sleeping radius, when the received SINR of the offloaded UEs is relatively low and the sleeping ratio increases slowly with available spectrum resource (SINR-limited);
    (2) low sleeping ratio with small sleeping radius, when the received SINR is high and the sleeping ratio mainly depends on the available spectrum resource (bandwidth-limited).
    When $\lambda_\mathrm{m}$ is high, increasing bandwidth for the offloaded UEs can significantly improve the sleeping ratio as shown in Fig.~\ref{fig_repulsive}(b).

    \emph{\textbf{(2) BS density}}

    As the black solid curves with triangles are completely overlapped by the red dashed curves in Figs.~\ref{fig_rand} and \ref{fig_repulsive}, deploying denser SCs does not improve network performance (consistent with Proposition~2).

    \emph{\textbf{(3) Influence of CB}}

    The gain brought by CB is marked in Figs.~\ref{fig_rand} and \ref{fig_repulsive}.
    Fig.~\ref{fig_rand}(a) and Fig.~\ref{fig_repulsive}(a) both reflect that CB brings higher performance gain when $\lambda_\mathrm{s}$ is small, which validates Proposition~3.
    In addition, CB can greatly improve the sleeping ratio under the repulsive scheme when the performance is bandwidth-limited (high $\lambda_\mathrm{m}$), as shown in Fig.~\ref{fig_repulsive}(b).\\

\begin{figure}
  \centering
  \includegraphics[width=2.5in]{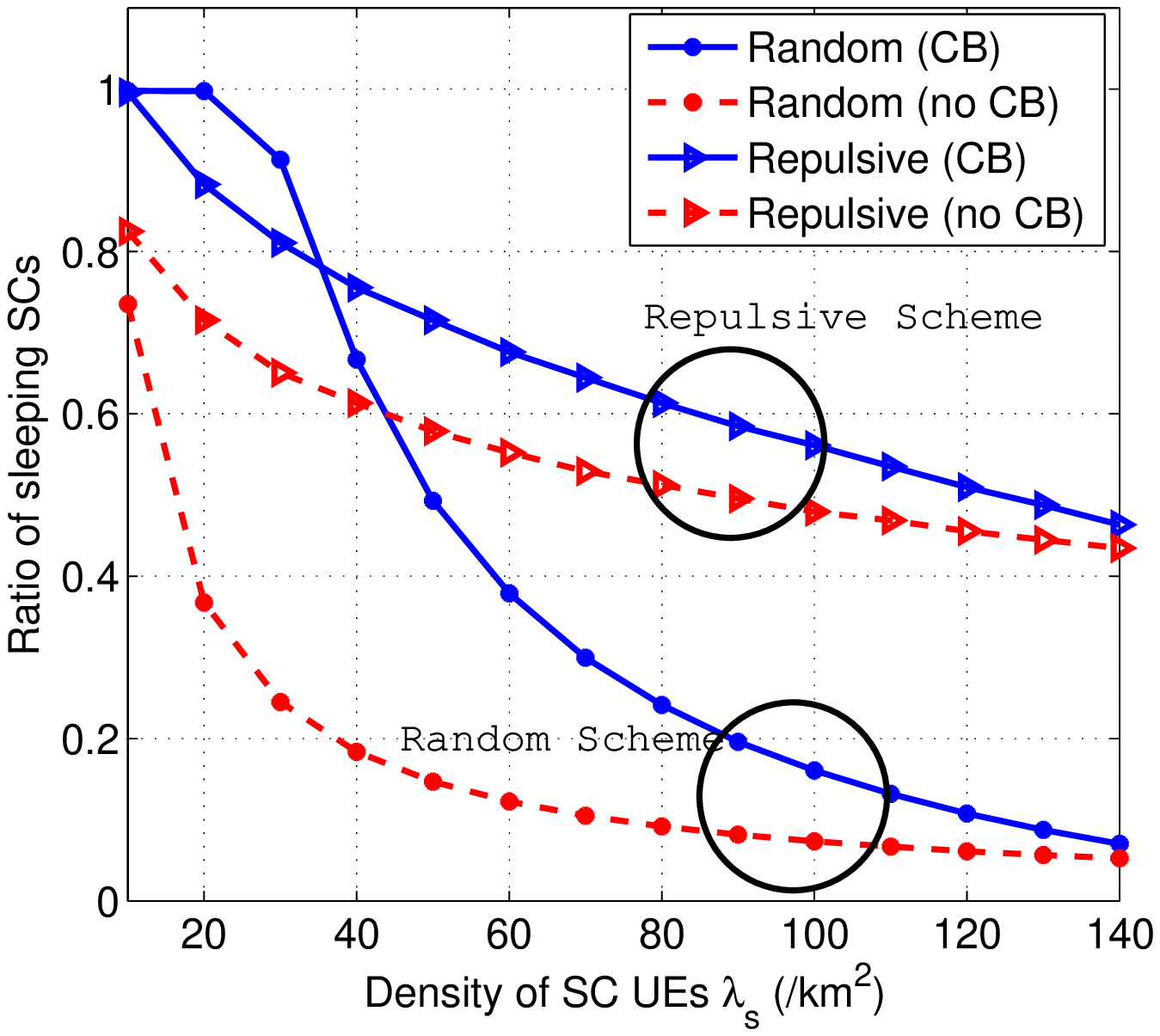}\\
  \caption{Comparison of two schemes ($\lambda_\mathrm{m}=200$/km$^2$).}
  \label{fig_compare}
\end{figure}

    Furthermore, the results of the two schemes under the baseline BS density are compared as shown in Fig.~\ref{fig_compare}.
    Generally, the sleeping ratio decreases more slowly with $\lambda_\mathrm{s}$ under the repulsive scheme, which is consistent with Proposition~5.
    When CB is not supported, repulsive scheme performs better than the random scheme.
    However, the random scheme is more advantageous than the repulsive scheme when $\lambda_\mathrm{m}$ is smaller and CB is conducted.
    This can be explained from two aspects: (1) The offloaded UEs enjoy higher spectrum efficiency under the repulsive scheme due to the smaller path loss, especially for the heavily loaded networks (Proposition~5); (2) Due to the lower inter-cell interference level, the SC-layer can provide more residual bandwidth for the offloaded UEs through CB under the random scheme, especially when the network is lightly-loaded and more SCs are turned off.\\

    Based on the above analysis and numerical results, we summarize our findings as follows.
    \begin{enumerate}
      \item Under the random scheme, the ratio of sleeping SCs is inversely proportional to $\lambda_\mathrm{s}$ while decreases linearly with $\lambda_\mathrm{m}$;
      \item Without CB, the repulsive scheme performs better;
      \item If CB is conducted, repulsive scheme performs better when the traffic load is high, otherwise the random scheme is better;
      \item Deploying more MBSs can help to improve network energy efficiency.
      \item CB brings higher performance gain for the networks with heavily-loaded MBS-layer and lightly-loaded SC-layer.
    \end{enumerate}

%% file: TrafficProfiles.tex
    \begin{figure}[!t]
        \centering
        \includegraphics[width=2.5in]{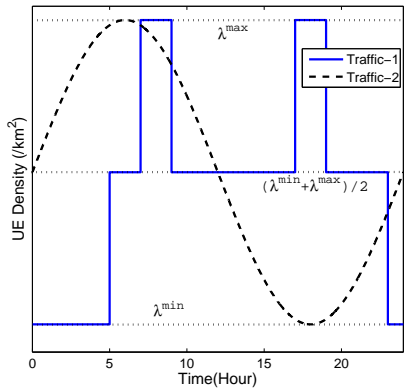}
        \caption{Daily traffic profiles.}
        \label{fig_traffic}
    \end{figure}

    \begin{figure}[!t]
        \centering
        \includegraphics[width=2.5in]{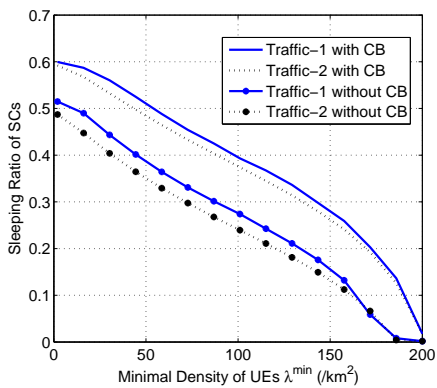}
        \caption{Average sleeping ratio of SCs.}
        \label{fig_energy_saving}
    \end{figure}
    
    In this part, we evaluate the performance of the two SC sleeping schemes under daily traffic profiles.
    Two typical traffic patterns are considered as shown in Fig.~\ref{fig_traffic}, where the x-axis denotes time and y-axis denotes the density of active UEs.
    We assume 80\% UEs require high data rate service, while the others are served at low data rate.
    Traffic pattern 1 describes the daily traffic variations of places like bus stations, whose two peaks correspond to the rush hours.
    Besides, traffic pattern 2 with sine function is often used to evaluate the effectiveness of energy saving algorithms \cite{Yiqun_Wu_ICCC} \cite{mine_IEICE}, \cite{Traffic_sine_1} and \cite{Traffic_sine_2}.

    For any given traffic load $\lambda(t)$, the maximum ratio of sleeping SCs under the two schemes can be obtained by solving Problems~(\ref{eq_formulation_random}) and (\ref{eq_formulation_repulsive}), and we dynamically choose the better one which turns off more SCs.
    The average ratio of sleeping SCs is shown in Fig.~\ref{fig_energy_saving} with parameters of Table~\ref{tab_parameter}.
    The peak traffic load of both traffic profiles is set as $\lambda^{\max} = 200$/km$^2$ (equivalent to the network capacity), whereas the minimal traffic load $\lambda^{\min}$ is set as a varying parameter reflecting the non-uniformity of the traffic load in time domain.
    Notice that different values of $\lambda^{\min}$ reflect different traffic non-uniformity.

    The ratio of sleeping SCs decreases with $\lambda_{\min}$, which indicates traffic dynamics help to turn off more SCs.
    As the network power consumption increases linearly with the density of active SCs without power control, this also suggests that networks with more fluctuated traffic can save more energy through SC sleeping.
    For the real systems, the traffic load after mid-night usually goes to zero.
    In this case, about 50\% SCs can be turned off on average without CB, and 10\% more SCs can go into sleep if CB is conducted.
    Besides, more SCs are expected to be turned off if effective SC sleeping algorithms are designed.
    Hence, HCN is an energy-efficient network architecture, under which on-demand service is provided through flexible SC sleeping and traffic offloading. 
    
    \begin{figure}[!t]
        \centering
        \includegraphics[width=2.5in]{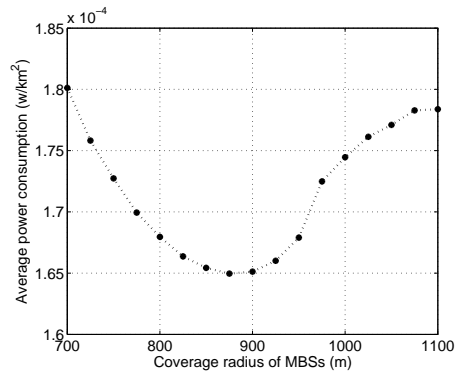}
        \caption{Network power consumption for different MBS density.}
        \label{fig_planning}
    \end{figure}

    Furthermore, we consider a network planning case when the maximal and minimal user densities are $\lambda^{\max}=$50/km$^2$ and $\lambda^{\min}=$0.5/km$^2$.
    To guarantee the basic coverage and service, the coverage radius of the MBSs should be no larger than 1100m. 
    Fig.~\ref{fig_planning} demonstrates the average power consumption per unit area for different MBS coverage radius, under traffic pattern 2. 
    It can be found that the energy-optimal coverage radius of the MBSs is 900m instead of 1100m, which indicates that deploying more MBSs can help to save energy.
    Although deploying denser MBSs causes higher energy consumption of the MBS-layer, more SCs can be turned off via vertical offloading, which helps to reduce energy consumption of the SC-layer.
    Thus there exists a tradeoff relation between the energy consumption of the two layers, and the energy-optimal MBS density may be higher than its minimal value required for basic coverage.
    The problem of energy-optimal network planning is left for our future work, due to the space limitations.

%% file: appendix.tex
\section{Proof of Theorem~1}
    \label{appendix_CBS}

    The inter-cell interference varies with the UE location, which makes the outage probability unsolvable.
    To derive closed-form expression, we average the inter-cell interference of all UE locations to approximate the uncertain value.
    Consider a MBS$_i$, the average inter-cell interference $\mathcal{\bar{I}}$ is
    \begin{equation}\label{eq_appendix_I}
        \bar{\mathcal{I}} = \int_{a\in \mathcal{A}_i} \sum\limits_{j\in \mathcal{B}^\mathrm{m},j \neq i} P^\mathrm{m}_{j} d_{aj}^{-\alpha_\mathrm{m}} f_{\mathcal{A}}(a) \mathrm{d} a,
    \end{equation}
    where $\mathcal{A}_i$ is the coverage area of MBS$_i$, and $a$ is the locations of UEs.
    Denote by $I=\bar{\mathcal{I}}/\sigma^2$ the ratio of inter-cell interference to noise for simplicity.

    Assume the number of residual UEs in the target MBS cell is given as $N_\mathrm{m}$,  then the probability that the data rate requirement of UE$_u$ can be satisfied is given by
        \begin{subequations} \footnotesize
            \label{eq_appendix_MBS_outage_Nm}
            \begin{align}
                & \mathds{P} \left\{\gamma_{u}^\mathrm{m} \geq 2^{(N_\mathrm{m}+1)\frac{U_\mathrm{m}}{ w_\mathrm{m}}} - 1 \right\} \nonumber \\
                &= \int_0^{D} \mathds{P} \left\{ h_u^\mathrm{m} \geq \frac{(I+1)\sigma^2}{P^{m}}d^{\alpha_\mathrm{m}} \left(2^{(N_\mathrm{m}+1) \frac{U_\mathrm{m}}{w_\mathrm{m}}} -1 \right)\right\} \frac{2d}{D^2} \mathrm{d} d  \nonumber \\
                &= \int_0^{D} \exp\left( -\frac{(I+1)\sigma^2}{P^{m}}d^{\alpha_\mathrm{m}} \left(2^{(N_\mathrm{m}+1) \frac{U_\mathrm{m}}{w_\mathrm{m}}} -1 \right)\right) \frac{2d}{D^2} \mathrm{d} d
                \label{eq_appendix_MBS_rayleigh} \\
                &= \int_0^{D} \left( 1-\frac{(I+1)\sigma^2}{P^{m}}d^{\alpha_\mathrm{m}} \left(2^{(N_\mathrm{m}+1) \frac{U_\mathrm{m}}{w_\mathrm{m}}} -1 \right)\right) \frac{2d}{D^2} \mathrm{d} d  \label{eq_appendix_MBS_PDF_rate} \\
                &= 1 - \frac{2D^{\alpha_\mathrm{m}}}{\alpha_\mathrm{m}+2} \frac{(I+1)\sigma^2}{P^{\mathrm{m}}} \left( 2^{(N_\mathrm{m}+1) \frac{U_\mathrm{m}}{w_\mathrm{m}}}-1 \right), \nonumber
            \end{align}
        \end{subequations}
    where MBS UEs are approximated to be uniformly distributed within a circle of radius $D$.
    Eq.~(\ref{eq_appendix_MBS_rayleigh}) holds as $h^\mathrm{m}_u$ follows exponential distribution, and (\ref{eq_appendix_MBS_PDF_rate}) is due to the assumption $\frac{\sigma^2}{P_\mathrm{m}}\rightarrow 0$. Although (\ref{eq_appendix_MBS_PDF_rate}) does not hold for the strong interference case (i.e., $I\rightarrow \infty$), (\ref{eq_appendix_MBS_PDF_rate}) applies to the MBS-layer, where most MBS UEs can receive high signal to interference ratio due to the large coverage radius.

    Recall that the probability distribution of $N_\mathrm{m}$ follows Poisson distribution.
    By substituting Eq.~(\ref{eq_appendix_MBS_outage_Nm}) into Eq.~(\ref{eq_outage_MBS}), the outage probability of a typical MBS UE is:
        \begin{equation} \footnotesize
            \begin{split}
                & 1 - G_\mathrm{m} = \sum_{N=0}^\infty \mathds{P} \left( \gamma_{u}^\mathrm{m} \geq 2^{(N+1) \frac{U_\mathrm{m}}{w_\mathrm{m}}} -1 \right) P_{N_\mathrm{m}}(N) \\
                &= \sum_{N=0}^\infty \mathds{P} \left( \gamma_{u}^\mathrm{m} \geq 2^{(N+1) \frac{U_\mathrm{m}}{w_\mathrm{m}}} -1 \right) \frac{(\frac{3\sqrt{3}}{2} D^2 \lambda_\mathrm{m})^N}{N!}e^{- \frac{3\sqrt{3}}{2} D^2 \lambda_\mathrm{m}} \\
                &=1-\frac{2 D^{\alpha_\mathrm{m}} (I\!+\!1)\sigma^2}{P^\mathrm{m} \left(\alpha_\mathrm{m}\!+\!2\right) } \! \left( 2^{\frac{U_\mathrm{m}}{w_\mathrm{m}}}\exp\left(\frac{3\sqrt{3}}{2} D^2 \lambda_{\mathrm{m}} \left( 2^{\frac{U_\mathrm{m}}{w_\mathrm{m}}} \!-\! 1 \right)\right)-1\right).
            \end{split}
        \end{equation}
    Hence, Theorem~1 is proved.

\section{Proof of Theorem~2}
    \label{appendix_TBS}

    When no SCs go into sleep, the distribution of the received SINR of a typical SC UE$_u$ in the interference-limited networks is given by \cite{JAndrews}:

    \begin{subequations}  \label{eq_appendix_PDF_SINR_PBS}\footnotesize
        \begin{align}
            & \int\limits_{0}\limits^{\infty} \mathds{P}\{\bar{\gamma}^\mathrm{s}_{u} \geq T \} f_{d_\mathrm{s}}(d) \mathrm{d} d\nonumber \\
            & \approx \frac{1}{1 + T^{\frac{2}{\alpha_\mathrm{s}}}\int_{T^{-\frac{2}{\alpha_\mathrm{s}}}}^{\infty} \frac{1}{1+x^{\frac{\alpha_\mathrm{s}}{2}}}\mathrm{d}x} \nonumber \\
            & \geq \frac{1}{1+T^{\frac{2}{\alpha_\mathrm{s}}}\int_{T^{-\frac{2}{\alpha_\mathrm{s}}}}^\infty x^{-\alpha/2} \mathrm{d} x} \label{eq_appendix_PDF_SINR_b} \\
            &= \frac{1}{1+\frac{2}{\alpha_\mathrm{s}-2}T}. \nonumber
        \end{align}
    \end{subequations}
    and the equality of (\ref{eq_appendix_PDF_SINR_b}) holds when $T \rightarrow 0$.

    Therefore, the probability that the data rate requirement of UE$_u$ can be satisfied is given by

    \begin{subequations}\footnotesize
        \label{eq_appendix_outage_2}
        \begin{align}
            & \mathds{P}\left\{\gamma^\mathrm{s}_{u} \geq 2^{ (N_\mathrm{s}+1) \frac{U_\mathrm{s}}{w_\mathrm{s}} } - 1 \right\} \nonumber\\
            & \approx \left( 1+\frac{2}{\alpha_\mathrm{s}-2} \left( 2^{ (N_\mathrm{s}+1) \frac{U_\mathrm{s}}{w_\mathrm{s}} } - 1 \right) \right)^{-1}
            \label{eq_appendix_outage_2_a} \\
            & = \frac{\frac{\alpha_\mathrm{s}-2}{2} 2^{ -(N_\mathrm{s}+1) \frac{U_\mathrm{s}}{w_\mathrm{s}} } } {\left( \frac{\alpha_\mathrm{s}-2}{2} -1 \right) 2^{ -(N_\mathrm{s}+1) \frac{U_\mathrm{s}}{w_\mathrm{s}} } + 1 } \nonumber\\
            & = \frac{(\alpha_\mathrm{s}-2)}{2}2^{ -\frac{U_\mathrm{s}}{w_\mathrm{s}} } \sum \limits_{m=0} \limits^{\infty} \left( \frac{4-\alpha_\mathrm{s}}{2} \right)^m 2^{-m}2^{ -m\frac{U_\mathrm{s}}{w_\mathrm{s}}} 2^{-(m+1) N_\mathrm{s} \frac{U_\mathrm{s}}{w_\mathrm{s}} } \label{eq_appendix_outage_2_b}
        \end{align}
    \end{subequations}
    where Eq.~(\ref{eq_appendix_outage_2_a}) holds for the assumption $\frac{U_\mathrm{s}}{w_\mathrm{s}} \rightarrow 0$.
    As the path loss factor generally satisfies $\alpha_\mathrm{s} \in (2,4]$, $1-\frac{\alpha_\mathrm{s}-2}{2} \in [0,1)$ and Eq.~(\ref{eq_appendix_outage_2_b}) holds.

    Furthermore, as $N_\mathrm{s}$ follows Poisson distribution with parameter $\lambda_\mathrm{s} A_\mathrm{s}$, we have
        \begin{equation}\footnotesize
            \begin{split}
                & \sum_{N=0}^{\infty} \mathds{P} \left\{ \gamma_{u}^\mathrm{s} \geq 2^{(N+1)\frac{U_\mathrm{s}}{w_\mathrm{s}}}-1 \right\} P_{N_\mathrm{s}}(N)\\
                = & \sum_{N=0}^{\infty} \frac{(\lambda_\mathrm{s} A_\mathrm{s})^N} {N!} \exp\{-\lambda_\mathrm{s} A_\mathrm{s}\} \mathds{P} \left\{ \gamma_{u}^\mathrm{s} \geq 2^{(N+1)\frac{U_\mathrm{s}}{w_\mathrm{s}}}-1 \right\} \\
                \approx & \frac{(\alpha_\mathrm{s}\!-\!2)}{2} 2^{ \!-\!\frac{U_\mathrm{s}}{w_\mathrm{s}} } \sum \limits_{m=0} \limits^{\infty} \left( \frac{4\!-\!\alpha_\mathrm{s}}{2} 2^{ - \! \frac{U_\mathrm{s}}{w_\mathrm{s}} } \right)^m \exp\left\{\lambda_\mathrm{s} A_\mathrm{s} \left( 2^{ -(m+1) \frac{U_\mathrm{s}}{w_\mathrm{s}} }\!-\!1\right) \right\}.
            \end{split}
        \end{equation}

    Recall that $A_\mathrm{s}$ follows Gamma distribution with shape $K$ and scale $\frac{1}{K \rho_\mathrm{s}}$:
        \begin{equation}\label{ }\footnotesize
            f_{A_\mathrm{s}}(A) = A^{K-1} \exp\{ -K \rho_\mathrm{s} A \} \rho_\mathrm{s}^K \frac{K^K}{\Gamma(K)},
        \end{equation}
    then we have
        \begin{equation}\footnotesize
            \label{eq_appendix_1_G}
            \begin{split}
            & \int_0^\infty \sum_{N=0}^{\infty} \mathds{P}\left\{ \gamma_{u}^\mathrm{s} \geq 2^{(N+1)\frac{U_\mathrm{s}}{w_\mathrm{s}}} -1 \right\}P_{N_\mathrm{s}}(N)f_{A_\mathrm{s}}(A)\mathrm{d}A \\
            = & \frac{(\alpha_\mathrm{s}-2)}{2}2^{ -\frac{U_\mathrm{T}}{w_\mathrm{T}}} \sum \limits_{m=0} \limits^{\infty}  \left( \frac{4\!-\!\alpha_\mathrm{T}}{2} 2^{ -\frac{U_\mathrm{s}}{w_\mathrm{s}} } \right)^m \left( \frac{\rho_\mathrm{s} K}{\rho_\mathrm{s} K +(1\!-\!2^{-(m+1)\frac{U_\mathrm{s}}{w_\mathrm{s}}})\lambda_\mathrm{s}} \right)^K.
            \end{split}
        \end{equation}

    Finally, due to the well-known exponential limit,
        \begin{eqnarray}\footnotesize
            \label{eq_appendix_exp_limit}
           & &  \lim_{\frac{U_\mathrm{s}}{w_\mathrm{s}} \rightarrow 0} \left\{ \frac{\rho_\mathrm{s} K}{\rho_\mathrm{s} K + \left( 1- 2^{-(m+1) \frac{U_\mathrm{s}}{w_\mathrm{s}}} \right)\lambda_\mathrm{s}} \right\}^K \nonumber \\
           &=& \lim_{\frac{U_\mathrm{s}}{w_\mathrm{s}} \rightarrow 0} \left\{1+\frac{1}{\frac{1}{\left( 1- 2^{-(m+1) \frac{U_\mathrm{s}}{w_\mathrm{s}}} \right)\lambda_\mathrm{s}}\rho_\mathrm{s} K }\right\}^{-K} \nonumber\\
           &=& \lim_{\frac{U_\mathrm{s}}{w_\mathrm{s}} \rightarrow 0} \exp \left\{ - \frac{\left( 1- 2^{-(m+1) \frac{U_\mathrm{s}}{w_\mathrm{s}}} \right)\lambda_\mathrm{s}}{\rho_\mathrm{s}} \right\} \\
           &=& \exp\left\{ -(m\! +\! 1)\frac{U_\mathrm{s}}{w_\mathrm{s}} \frac{\lambda_\mathrm{s}}{\rho_\mathrm{s}} \log2 \right\} = 2^{-(m+1) \frac{U_\mathrm{s}}{w_\mathrm{s}} \frac{\lambda_\mathrm{s}}{\rho_\mathrm{s}}}.\nonumber
        \end{eqnarray}
    By substituting (\ref{eq_appendix_exp_limit}) into (\ref{eq_appendix_1_G}), we have
        \begin{equation} \footnotesize
            \begin{split}
               & \int_0^\infty \sum_{N=0}^{\infty} \mathds{P}\left\{ \gamma_{u}^\mathrm{s} \geq 2^{(N+1)\frac{U_\mathrm{s}}{w_\mathrm{s}}} -1 \right\}P_{N\mathrm{s}}(N)f_{A_\mathrm{s}}(A)\mathrm{d}A \\
               \approx & \frac{(\alpha_\mathrm{s}-2)}{2} 2^{ -\frac{U_\mathrm{s}}{w_\mathrm{s}} (1+\frac{\lambda_\mathrm{s}}{\rho_\mathrm{s}}) }\sum \limits_{m=0} \limits^{\infty} \left\{ \left( \frac{\alpha_\mathrm{s}-2}{2} \!-\!1 \right) 2^{-\frac{U_\mathrm{s}}{w_\mathrm{s} } (1+\frac{\lambda_\mathrm{s}}{\rho_\mathrm{s}})} \right\}^m \\
               = & \frac{\frac{\alpha_\mathrm{s}-2}{2} 2^ {-\frac{U_\mathrm{s}}{w_\mathrm{s}} (1+\frac{\lambda_\mathrm{s}}{\rho_\mathrm{s}})}}{ 1 - \frac{4-\alpha_\mathrm{s}}{2} 2^{-\frac{U_\mathrm{s}}{w_\mathrm{s}}(1+\frac{\lambda_\mathrm{s}}{\rho_\mathrm{s}}) }  }.
            \end{split}
        \end{equation}
    Hence, Theorem~2 is proved.

%% file: manuscript.bbl
\begin{thebibliography}{100}

    \bibitem{JSAC_overview_5G} J.~Andrews, et al., ``What will 5G be?'', \emph{IEEE J. Sel. Areas Commun.}, vol.~32, no.~6, pp.~1065-1082, May, 2013.
    \bibitem{5G_backhaul} X.~Ge, H.~Cheng, M.~Guizani, T.~Han, ``5G wireless backhaul networks: challenges and research advances,'' \emph{IEEE Network}, vol.~28, no.~6, pp.~6-11, Nov., 2014.
    \bibitem{NZhang_cloud_5G} N.~Zhang, N.~Cheng, A.~Gamage, K.~Zhang, J.~Mark, and X.~Shen, "Cloud assisted HetNets toward 5G wireless networks", \emph{IEEE Commun. Magazine}, to appear.
    \bibitem{Overview_green} L.~Suarez, L.~Nuaymi, and J.~Bonnin, ``An overview and classification of research approaches in green wireless networks,'' \emph{EURASIP J. Wireless Commun. and Netw., Special Issue: Green Radio}, Apr., 2012.
    \bibitem{Green_MIMO} X.~Ge, X.~Huang, Y.~Wang, M.~Chen, Q.~Li, T.~Han and C.~Wang, ``Energy efficiency optimization for MIMO-OFDM mobile multimedia communication systems with QoS constraints,'' \emph{IEEE Transactions on Vehicular Technology}, vol.~63, no.~5, pp.~2127-2138, June 2014.

    \bibitem{Tango} Z.~Niu, ``TANGO: traffic-aware network planning and green operation'', \emph{IEEE Wireless Commun.}, vol.~18, pp.~22-29, Oct. 2011.

    \bibitem{Hyper_cellular} Z.~Niu, S.~Zhou, S.~Zhou, X.~Zhong, and J.~Wang, ``Energy efficiency and resource optimized hyper-cellular mobile comminication system architecture and its technical challenges,'' \emph{Scientia Sinica(Informationis)}, vol.~42, no.~10, pp.~1191-1203, 2012.
    \bibitem{Beyond_green} A.~Capone, A.~F.~dos~Santos, I.~Filippini, and B.~Gloss, ``Looking beyond green cellular networks,'' in \emph{IEEE WONS'12}, Courmayeur, Italy, Jan. 2012.
    \bibitem{Docomo_phantom} H.~Ishii, Y.~Kishiyama, and H.~Takahashi, ``A novel architecture for LTE-B C-plane/U-plane split and phantom cell concept,'' in \emph{IEEE GLOBECOM'12}, Anaheim, CA, USA, pp. 624-630, 2012.
    \bibitem{3GPP_release12} 3GPP TR 36.842 V0.2.0, ``Study on small cell enhancements for E-UTRA and E-UTRAN - Higher layer aspects,'' 2013.
    \bibitem{Huawei} X.~Xu, G.~He, S.~Zhang, Y.~Chen, and S.~Xu, ``On functionality separation for green mobile networks: concept study over LTE,'' \emph{IEEE Commun. Mag.}, vol.~51, no.~5, pp.~82-90, 2013.
    \bibitem{5G_Datang} S.~Chen, J.~Zhao, ``The requirements, challenges, and technologies for 5G of terrestrial mobile telecommunication,'' \emph{IEEE Commun. Mag.}, vol.~52, no.~5, pp.~36-43, 2014.
    \bibitem{5G_magzine} B.~Bangerter, S.~Talwar, R.~Arefi, and K.~Stewart, ``Networks and devices for the 5G era,'' \emph{IEEE Commun. Mag.}, vol.~52, no.~5, pp.~90-96, 2014.

    \bibitem{EARTH} G.~Auer, et~al., ``D2.3: energy efficiency analysis of the reference systems, areas of improvements and target breakdown,'' INFSO-ICT-247733 EARTH, Tech. Rep., Nov., 2010. [Online]. Available: www.ict-earth.eu/publications/ deliverables/deliverables.html.

    \bibitem{Cell_zooming} Z.~Niu, Y.~Wu, J.~Gong, and Z.~Yang, "Cell zooming for cost-efficient green cellular networks," \emph{IEEE Commun. Mag.}, vol.~48, no.~11, pp.~74-79, Nov. 2010.
    \bibitem{SZhou_BS_sleeping} S.~Zhou, J.~Gong, Z.~Yang, Z.~Niu and P.~Yang, "Green mobile access network with dynamic base station energy saving," in \emph{ACM MobiCom'09}, Beijing, China, Sept. 2009.
    \bibitem{night-zone-optimal-ratio} M.~Marsan, L.~Chiaraviglio1, D.~Ciullo1, and M.~Meo, ``Optimal energy savings in cellular access networks,'' in \emph{IEEE ICC'09}, Dresden, Germany, June, 2009.
    \bibitem{JGong-Wiopt} J.~Gong, S.~Zhou, Z.~Niu, and Y.~Peng, ``Traffic-aware base station sleeping in dense cellular networks,'' in \emph{IEEE WiOpt'10}, Avignon, France, June, 2010.
    \bibitem{night-zone} L.~Chiaraviglio, D.~Ciullo, M.~Meo, and M.~Marsan, ``Energy-aware UMTS access networks,'' in \emph{WPMC'08}, Lapland, Finland, Sept. 2008.
    \bibitem{Yiqun_Wu_ICCC} Y.~Wu, and Z.~Niu, ``Energy efficient base station deployment in green cellular networks with traffic variations,'' in \emph{IEEE ICCC'12}, Beijing, China, Aug. 2012.
    \bibitem{Tony_single_tier_sleep} Y.~Soh, T.~Quek, M.~Kountouris, and H.~Shin, ``Energy efficient heterogeneous cellular networks,'' \emph{IEEE J. Sel. Areas Commun.}, vol.~31, no.~5, pp.~840-850, May, 2013.
    \bibitem{mine_IEICE} S.~Zhang, Y.~Wu, S.~Zhou, and Z.~Niu, ``Traffic-aware network planning and green operation with BS sleeping and cell zooming,'' in \emph{IEICE Trans. Commun.}  vol.~E97-B, no.~11, pp.~2337-2346, Nov. 2014.


    \bibitem{DCao} D.~Cao, S.~Zhou, Z.~Niu, ``Optimal combination of base station densities for energy-efficient two-tier heterogeneous cellular networks,'' \emph{IEEE Trans. Wireless Commun.}, vol.~12, no.~9, pp.~4350-4362, Sept. 2013.
    \bibitem{Repulsive} S.~Cho, and W.~Choi, ``Energy-efficient repulsive cell activation for heterogeneous cellular networks,'' \emph{IEEE J. Sel. Areas Commun.}, vol.~31, no.~5, pp.~870-882, May, 2013.

    \bibitem{forcast_exponential_smoothing} S.~Morosi, P.~Piunti, and E.~Re, ``Improving cellular network energy efficiency by joint management of sleep mode and transmission power,'' in \emph{IEEE TIWDC'13}, Genoa, Italy, Sept. 2013.
    \bibitem{forcast_holt-Winter} S.~Morosi, P.~Piunti, and E.~Re, ``Sleep mode management in cellular networks: a traffic based technique enabling energy saving, ''\emph{Trans. Emerging Tel. Tech.}, vol.~24, issue~3, pp.~331-341, 2013. [Online], Available: http://onlinelibrary.wiley.com/doi/10.1002/ett.2621/abstract
    \bibitem{forcast_ANN} G.~Wang, C.~Guo, S.~Wang, and C.~Feng, ``A traffic prediction based sleeping mechanism with low complexity in femtocell networks,'', in \emph{IEEE ICC'13}, Budapest, Hungary, June, 2013.

    \bibitem{vertical_MDP} L.~Saker, S.~Elayoubi, R.~Combes, T.~Chahed, ``Optimal control of wake up mechanisms of femtocells in heterogeneous networks,'' \emph{IEEE J. Sel. Areas Commun.}, vol.~30, no.~2, pp.~664-672, Apr., 2012.
    \bibitem{Mine_Globecom_load} S.~Zhang, J.~Wu, J.~Gong, S.~Zhou, and Z.~Niu, ``Energy-optimal probabilistic base station sleeping under a separation network architecture,'' in \emph{IEEE GLOBECOM'14}, Austin, USA, Dec., 2014.

    %\bibitem{night_zone_two_networks} M.~Marsan, and M.~Meo, ``Energy efficient management of two cellular access networks,'' \emph{ACM SIGMETRICS Performance Evaluation Review}, vol.~37 issue~4, 2010.

    \bibitem{WZhang_separation} Z.~Wang, and W.~Zhang, ``A separation architecture for achieving energy-efficient cellular networking,'' \emph{IEEE Trans. Wireless Commun.}, vol.~13, no.~6, pp.~3113-3123, June, 2014.
    \bibitem{mine_asilomar} S.~Zhang, S.~Zhou, and Z.~Niu, ``Traffic aware offloading for BS sleeping in heterogeneous networks,'' in \emph{Asilomar Conference on Signals, Systems, and Computers'15}, California, USA, Nov., 2014.

    %\bibitem{JWu_ICC_signaling_taffic} J.~Wu, S.~Zhou, Z.~Niu, C.~Liu, P.~Yang, and G.~Miao, ``Traffic-aware data and signaling resource management for green cellular networks,'' in \emph{IEEE ICC'14}, Sydney, Australia, 2014.

    %\bibitem{3GPP_Release12} 3rd Generation Partnership Project, ``Overview of 3GPP Release 12 V0.1.2,'' 2014. [Online]. Available: http://www.3gpp.org/ftp/ Information/WORK\_PLAN/Description\_Releases/. [Accessed: Mar. 25, 2014].

    \bibitem{JAndrews} J.~Andrews, F.~Baccelli, and R.~Ganti, ``A tractable approach to coverage and rate in cellular networks,'' \emph{IEEE Trans. Wireless Commun.}, vol.~51, no.~11, pp.~3122-3134, Nov. 2011.

    \bibitem{Stochastic_Geometry} S.~N.~Chiu, D.~Stoyan, W.~S.~Kendall, and J.~Mecke, \emph{Stochastic Geometry and Its Applications}, 3rd Edition, UK: John~Wiley~\&~Sons, 2013, pp.~48-51.

    \bibitem{Slivnyak_theorem} M.~Heanggi, J.~Andrews, F.~Baccelli, O.~Dousse, and M.~Franceschetti, ``Stochastic geometry and random graphs for the analysis and design of wireless networks,'' invited paper, \emph{IEEE J. Sel. Areas Commun.}, vol.~27, no.~7, pp.~1029-1046, Sept., 2009.

    \bibitem{Traffic_sine_1} M.~Marsan, L.~Chiaraviglio, D.~Ciullo, and M.~Meo, ``Energy-efficient managment of UMTS access networks,'' in \emph{IEEE ITC'09}, Paris, France, Sept., 2009.
    \bibitem{Traffic_sine_2} L.~Chiaraviglio, D.~Ciullo, G.~Koutitas, M.~Meo, and L.~Tassiulas, ``Energy-efficient planning and management of cellular networks,'' in \emph{IEEE WONS'12}, Courmayeur, Italy, Jan., 2012.





\end{thebibliography}
